\documentclass[12pt,a4paper]{article}

\usepackage{jheppub}

\makeatletter
\gdef\@fpheader{}
\makeatother

\usepackage[latin1]{inputenc}
\usepackage{amsmath}

\usepackage{amsfonts}
\usepackage{physics}
\usepackage{amssymb}
\usepackage{faktor}
\usepackage{graphicx}
\usepackage{tikz}
\usepackage{amsthm}
\numberwithin{equation}{section}
\usepackage[english]{babel}
\usepackage{mathtools}
\usepackage{nicefrac}
\usepackage{lscape}
\usepackage{textgreek}
\usepackage{parskip}
\usepackage{enumerate}

\usepackage{cleveref}

\usepackage{tcolorbox}

\usepackage{todonotes}

\usepackage{dsfont}

\usepackage{url}

\theoremstyle{definition}

\usepackage{subcaption}
 \usepackage{cleveref}
 \crefname{figure}{figure}{figures}
\Crefname{figure}{Figure}{Figures}

\hypersetup{
	pdftitle={Unorientable JT gravity},
	pdfauthor={Jarod Tall}}
\usepackage{upgreek}
\usepackage{array}
\usepackage{bm}

\author[a]{Jarod Tall,}
\author[b]{Torsten Weber,}
\author[b]{Juan Diego Urbina,}
\author[b]{and Klaus Richter}
\affiliation[s]{Department of Physics and Astronomy, Washington State University,\\
Pullman, WA 99164-2814 USA}
\affiliation[b]{Institut f\"ur Theoretische Physik, 
Universit\"at Regensburg, \\
Universit\"atsstr. 31, D-93040 Regensburg, Germany}

\emailAdd{jarod.tall@wsu.edu}
\emailAdd{torsten.weber@ur.de}
\emailAdd{juan-diego.urbina@ur.de}
\emailAdd{klaus.richter@ur.de}
\title{
Chaos and moduli space volumes in unorientable JT gravity}
\abstract{We show the late time, or $\tau-$scaled, limit of the canonical spectral form factor (SFF) in unorientable JT gravity agrees with universal random matrix theory (RMT) up to genus one in the topological expansion, establishing a key signature of quantum chaos for the time-reversal symmetric case. The loop equations for an orthogonal matrix model with spectral curve $y(z) \propto \sin(2\pi z)$ are used to compute the moduli space volumes of unorientable surfaces. The divergences of the unorientable volumes are regularized by first regularizing the resolvents of the orthogonal matrix model. To this end, we make use of the large $p$ limit of the $(2,2p+1)$ minimal string model. Using properties of the volumes and the loop equations, we derive streamlined formulas to compute the volumes for one and two boundaries, giving explicit results up to genus one. We find the general structure of the unorientable volumes to be written in terms of multiple polylogarithms and zeta values, with weight determined by the genus, number of boundaries, and number of crosscaps. In the $\tau-$scaled limit, contributions to the SFF from the divergent parts of the volume cancel, and the SFF becomes finite and independent of regularization. The SFF from universal RMT is a distinct computation, that depends on the leading order energy density of JT gravity, which we also derive up to genus one.  } 
\makeindex

\begin{document}
\maketitle
\section{Introduction}
Jackiw-Teitelboim (JT) gravity \cite{Jackiw1985,Teitelboim1983} provides a simple model to study two dimensional quantum gravity. The bulk theory describes two dimensional hyperbolic surfaces, to which nearly AdS$_2$ asymptotic boundaries are attached \cite{Saad2019}. The boundary mode is described by the Schwarzian action, well known to also describe the low energy limit of the Sachdev-Ye-Kitaev (SYK) model \cite{Kitaev2016_1,Kitaev2016_2, Maldacena2016b, Jensen2016}. A very nice review of JT gravity can be found in \cite{Mertens2023}. Although JT gravity is a relatively simple theory, it becomes very rich when all of the different variations of the bulk geometric theory are considered. In the work of Stanford and Witten each geometric variation was enumerated and shown to correspond to a choice of discrete symmetry \cite{Stanford2019}. The choice of symmetry then dictates a corresponding random matrix theory (RMT) ensemble, which can be one of the three Wigner-Dyson ensembles or the seven Altland-Zirnbauer ensembles \cite{Dyson1962a, Altland1997}. In this way, a duality between each variation of JT gravity and a matrix model was conjectured, something that was proven for the simplest case of JT gravity dual to a unitary matrix model in \cite{Saad2019}, and later proven for the orthogonal case in \cite{Stanford2023}.
To understand the different geometric variations of JT gravity it is useful to think of the boundary theory as being described by a quantum Hamiltonian, such as the SYK model, with a particular choice of discrete symmetry, then by invoking the NAdS$_2$/NCFT$_1$ dictionary \cite{Maldacena2016}, the choice of symmetry can be related to the topological properties of the bulk. For example, if the boundary theory Hamiltonian has time-reversal symmetry this translates to the inclusion of unorientable surfaces in the bulk theory. This correspondence then leads to the association of JT gravity with time-reversal symmetry as unorientable JT gravity. Time-reversal invariance corresponds to either an orthogonal or symplectic matrix model, the former of which will be the focus of the present paper.

The duality between JT gravity and a matrix model is interesting from the perspective of quantum chaos, since the spectral statistics of a quantum chaotic Hamiltonian can be described by RMT in the proper universal regime \cite{Bohigas1984}. The duality to a matrix model then suggests JT gravity behaves as a quantum chaotic system. The notion can be made concrete by computing late time correlation functions in JT gravity, such as the (canonical)
spectral form factor
(SFF)\footnote{In the following by SFF we always refer to the canonical spectral form factor.}, and comparing the results to the predictions of universal RMT \cite{Saad2018, Cotler2017, Altland2020}. The agreement for the late time limit of the SFF has already been established analytically for JT gravity without time reversal symmetry~\cite{Saad2022}. In the authors' prior work agreement was established for the low energy limit of unorientable JT gravity, i.e. unorientable topological gravity or the unorientable Airy model \cite{Weber2024}. Here we generalize much of the results of \cite{Weber2024} to the full theory of unorientable JT gravity. Extending the results to unorientable JT gravity encounters difficulty since the theory is formally divergent and requires regularization as we will explain momentarily.

\begin{figure}[h]
				\centering
				\begin{subfigure}[b]{0.4\textwidth}
					\centering
					\includegraphics[width=\textwidth]{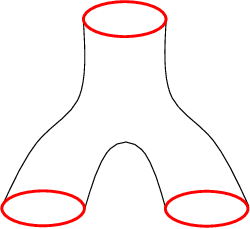}
					\caption{}
					\label{fig:PoP}
				\end{subfigure}
				~
				\begin{subfigure}[b]{0.4\textwidth}
					\centering
					\includegraphics[width=\textwidth]{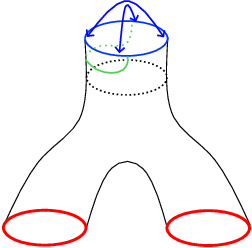}
					\caption{}
					\label{fig:PocP}
				\end{subfigure}
				\caption{The building blocks of hyperbolic surfaces: a) A three holed sphere with geodesic boundaries in red. b) A three holed sphere with a crosscap attached at one geodesic boundary (indicated by the broken line). The blue arrows indicate the identification of antipodal points on the blue boundary of the attached crosscap. The green curve is included as its neighborhood is homeomorphic to a M\"obius strip.}
			\end{figure}

Obtaining correlation functions in JT gravity reduces to computing the moduli space volume of hyperbolic surfaces of genus $g$ with $n$ geodesic boundaries of lengths $b_1, \dots, b_n$. The building blocks of an arbitrary hyperbolic orientable surface are hyperbolic three-holed spheres,
i.e. pairs of pants, pictured in \cref{fig:PoP}. To build an unorientable surface, characterized by containing a M\"obius strip, an additional building block known as a crosscap is needed. The easiest way to construct such a crosscap is by taking a cylinder and identifying antipodal points on one of its boundaries \cite{Stanford2019}, pictured in \cref{fig:PocP}. This geometry is unorientable as seen by inspecting the neighborhood of the green curve in \cref{fig:PocP} and recognizing it to be homeomorphic to a M\"obius strip. The set of fundamental building blocks is thus augmented by three-holed spheres with one or two glued crosscaps\footnote{A special case is given by a surface with $(g,n)=(\frac 12,1)$ which is the crosscap geometry itself. It should also be noted an unorientable surface can be created by gluing two circular boundaries with reversed orientations \cite{Norbury2008,Stanford2023}.}.

For the simplest case, where the surfaces are taken to be orientable, the moduli space volumes can be computed using Mirzakhani's recursion relation \cite{Mirzakhani2006, Mirzakhani2007}. The inclusion of unorientable surfaces leads to logarithmic divergences in the moduli space volumes, stemming from contributions of small crosscaps \cite{Norbury2008, Gendulphe2017}. Recently, Stanford generalized Mirzakhani's recursion to include unorientable surfaces by imposing a cut-off on the measure of crosscaps, which led to defining the regularized volumes \cite{Stanford2023}. 
Practical computations of the regularized volumes, necessary for correlation functions in unorientable JT gravity, appear out of reach due to the difficulty of the integrals involved in Stanford's recursion. For example, for the case of $(g,n)=(1,1)$ the finite part of the volume was only able to be computed numerically, though the case of $(g,n)=(\frac{1}{2},2)$ can be computed in full \cite{Saad2022, Stanford2023}.

The strategy taken in this paper is to compute the regularized volumes using the loop equations of the dual matrix model. 
It was originally shown by Eynard and Orantin that Mirzakhani's recursion relation, where only orientable surfaces are considered, maps directly to the loop equations of a double scaled matrix model with unitary symmetry class and spectral curve\footnote{The spectral curve is directly related to the leading order density of eigenvalues of the matrix model, i.e. $\rho_0(E) \propto \sinh(2\pi \sqrt{E})$ with $E=-z^2$} $y(z) \propto \sin(2\pi z)$ \cite{Eynard2007}. The relationship between Mirzakhani's recursion and the loop equations, is at the core of the duality between JT gravity and a matrix integral that was established in \cite{Saad2019}. The case of bosonic JT gravity with the inclusion of unorientable surfaces has two sub-cases dual to matrix models with either orthogonal or symplectic symmetry. The duality conjectured in \cite{Stanford2019} for these cases was proven in \cite{Stanford2023} where it was shown the recursion for the unorientable volumes maps directly to the loop equations of an orthogonal matrix model. The symplectic case would follow similarly by including a negative sign for each crosscap as explained in \cite{Stanford2019}. In this paper we only consider the orthogonal case, and the term unorientable JT gravity will be reserved for 
this case only.

In the unitary/orientable case the loop equations reduce to topological recursion, which allows the orientable volumes to be found with a simple residue computation. The important point is computing the necessary residues in topological recursion is easier than computing the integrals involved in Mirzakhani's recursion. Though in the orthogonal case the loop equations do not reduce to topological recursion, an analogous situation occurs. We show the loop equations for an orthogonal matrix model with the JT gravity spectral curve reduces to computing residues,  which are somewhat more favorable than the integrals involved in Stanford's recursion. However, unlike the unitary case, there are now infinitely many residues that have to be computed. Fortunately, it turns out that many of the residues do not contribute to the unorientable volumes and thus do not need to be computed in the first place.  
This simplification leads to streamlined formulas that greatly reduce computational effort, allowing the computation of the entire $(g,n)=(1,1)$ and $(g,n) = (1,2)$ unorientable volumes. We find the unorientable volumes to have a multiple polylogarithmic structure that generalizes the structure of the orientable volumes. Due to this structure and algebraic properties of multiple polylogarithms, the unorientable volumes are surprisingly simple compared to what would be initially expected.

The central aim of this work is to compute the late time, or $\tau$-scaled \cite{Saad2022, Weber2022, Blommaert2022,Okuyama2021, Okuyama2023, Weber2024}, limit of the SFF from unorientable JT gravity in two distinct ways and show they are equivalent to give a clear indication for the presence of quantum chaos in unorientable JT gravity. The first method computes the $\tau-$scaled SFF directly from the unorientable volumes up to $g=1$ in the genus expansion, which corresponds to the cubic term of rescaled time. The second method computes the $\tau-$scaled SFF from universal RMT in the Gaussian orthogonal ensemble (GOE) by taking the Laplace transform of the \textit{microcanonical} SFF. The universal form of the microcanonical SFF is well established in the quantum chaos literature \cite{Mehta2004, Haake2010}\footnote{The microcanonical SFF is defined in terms of the form factor which is given in \cite{Mehta2004, Haake2010}.}, and only depends on the symmetry class and the leading order density of states. It should be emphasized that, though the universal form of the \textit{microcanonical} SFF is known, finding the \textit{canonical} SFF, i.e. the $\tau-$scaled SFF, from the \textit{microcanonical} SFF  involves a non-trivial computation for the JT gravity case. We succeed in deriving the universal RMT result up to $g=1$, allowing the comparison to the gravity results, and introduce a method that can be extended to higher genus if desired. We show that these two distinct computations of the SFF match up to a subtlety in the Airy terms that was discussed in \cite{Weber2024}.

The structure of the paper is as follows:  In section \ref{background} we briefly review background material on how the the moduli space volumes can be used to compute correlation functions in JT gravity, introduce the regularized unorientable volumes, the corresponding regularized resolvents and our regularization procedure, which differs from Stanford's. Having done this, we summarize the main results of the paper in section \ref{summary}. In section \ref{section2}, we introduce the loop equations and show how to compute the regularized resolvents. From the resolvents, we show how to compute the regularized volumes and introduce a streamlining procedure that simplifies computations. In section \ref{section3}, we first use the unorientable volumes to compute the long time limit of the SFF in unorientable JT gravity (\cref{section3a}). We then compute the analogous result from universal RMT (\cref{section3b}) and compare. In the appendices we go through the details of computing the unorientable volumes (\cref{details}), the SFF in JT gravity (\cref{appendixc}), and the SFF from universal RMT (\cref{appendixd}).

\subsection{Background}
\label{background}
We are, motivated by the canonical SFF, interested in connected correlation functions of partition functions in JT gravity and their topological expansion. 
The connected two-point function in JT gravity is found to have the following topological expansion
\cite{Stanford2019}:
\begin{align}
  \ev{Z(\beta_1)Z(\beta_2)}_c \cong \sum_{g=0,\frac{1}{2},1 \dots}^{\infty}\frac{Z_{g}(\beta_1,\beta_2)}{\left(e^{S_0}\right)^{2g}},
\end{align}
where $\cong$ signifies an asymptotic series and 
\begin{align}
\label{p}
  Z_{g}(\beta_1,\beta_2)=\int_0^\infty \int_0^\infty b_1\dd{b_1} b_2\dd{b_2}Z^t(b_1,\beta_1) Z^t(b_2,\beta_2)V_{g}(b_1,b_2),
\end{align}
with the trumpet partition function defined as
\begin{align}
  Z^t(b,\beta)\coloneqq\frac{1}{\sqrt{4\pi \beta}}e^{-\frac{b^2}{4\beta}}.
\end{align}
Furthermore, $V_g(b_1,b_2)$ are the moduli space volumes of unorientable surfaces of constant negative curvature with genus $g$ and $n=2$ geodesic boundaries of lengths $b_1$ and $b_2$. This construction can be directly generalized to an arbitrary number of boundaries. As discussed in the introduction, the unorientable volumes are logarithmically divergent and require regularization. In \cite{Stanford2023}, Stanford regularized the volumes by imposing a cut-off on the measure of crosscaps, leading to the following expansion of the $p$-regularized volumes:\footnote{To match the definition given in \cite{Stanford2023} take $p \propto 1 /\epsilon$ .}
\begin{align}
  V_g^{(p)}(b_1,\dots,b_n) = \sum_{k=0}^{2g}\log(p)^{k}v_{g,k}(b_1,\dots,b_n)+\order{p^{-1}}.
  \label{eq:regvol}
\end{align}
Throughout the paper we will sometimes refer to the coefficients, $v_{g,k}(b_1, \dots, b_n)$, as the unorientable volumes. We can think of these coefficients as the moduli space volumes of surfaces with genus $g$, $n$ boundaries, and $k$ crosscap divergences.
The regularized volumes are what will be used to compute the two point function, so the replacement $V_g(b_1,b_2) \to V^{(p)}_g(b_1,b_2) $ should be made in~\eqref{p}. To compute the unorientable volumes we will use the relationship to the correlation functions of resolvents of a double scaled matrix model in the orthogonal symmetry class.
(Multi-)resolvents are defined as:
 \begin{align}
  R(x_1,\ldots,x_n)=\Tr\frac{1}{x_1-H}\ldots\Tr\frac{1}{x_n-H}.
\end{align} 
The connected correlation functions of resolvents in a double scaled matrix model can be written as a pertubative expansion in the parameter $e^{S_0}$:\footnote{Here we use the standard notion of a double scaled matrix model where the parameter $e^{S_0}$ governs the perturbative expansion of correlation functions, see for instance \cite{Saad2019}. For an in depth treatment of double scaling and genus expansions see chapter 5 of \cite{Eynard2016}.}
\begin{align}
  \expval{R(x_1,\dots,x_n)}_c \cong \sum_{g=0,\frac{1}{2},1, \cdots} \frac{R_g(x_1,\dots,x_n)}{e^{S_0(2g+n-2)}}.
\end{align}
If we specify the matrix model to be in the orthogonal symmetry class and the spectral curve to be
\begin{align}
y(z) = \frac{\sin(2\pi z)}{4\pi},
\label{eq:sinh}
\end{align}
the relationship to the $p-$regularized unorientable volumes is given by the inverse Laplace transform
\cite{Stanford2023}, which is a direct generalization of the orientable case \cite{Eynard2007}:
\begin{equation}
    \begin{aligned}
        V^{(p)}_{g}(b_1,\dots,b_n)&=\mathcal{L}^{-1}\qty[\prod_{i=1}^n\qty( \frac{-2z_i}{b_i})R^{(p)}_{g}(-z_1^2,\dots,-z_n^2),\qty(b_1,\dots,b_n)] \\
  &=(-1)^n\int_{\delta +i\mathbb{R}}R^{(p)}_{g}\left(-z_1^2,\dots,-z_n^2\right)\prod_{j=1}^n \frac{dz_j}{2\pi i}\frac{2z_j}{b_j}e^{b_j z_j}. 
  \label{eq:volume}
    \end{aligned}
\end{equation}
Since the volumes are divergent, this relationship implies that the contributions to the correlation functions of resolvents on the right hand side are in need of regularization as well. Consequently, we need to regularize the resolvents in a way that corresponds to~\eqref{eq:regvol}. The regularization can be accomplished by defining the $p$-regularized spectral curve, $y^{(p)}(z)$, in the following way:
\begin{align}
  \lim_{p \to \infty} y^{(p)} (z) = \frac{\sin(2\pi z)}{4\pi}.
\end{align}
It is also important for $y^{(p)} (z)$ to reproduce the Airy model spectral curve at $p=0$, i.e. $y^{\text{Airy}}(z)=\frac{z}{2}$. The spectral curve of the $(2p+1)$ minimal string model has these exact properties, as well as a known gravitational dual. Thus, as it was pointed out in Appendix F of \cite{Stanford2019}, the $(2,2p+1)$ minimal string model can be used to regularize unorientable JT gravity. The spectral curve is \cite{Artemev2024,Mertens2021}
\begin{align}
  y^{(p)}(z) = (-1)^p\frac{1}{4\pi}T_{2p+1}\left(\frac{2\pi z}{2p+1}\right),
  \label{eq:yp}
\end{align}
where $T_{2p+1}(x)$ is the Chebyshev polynomial of the first kind. The physical interpretation of the regularization of the spectral curve is that we are imposing a high energy cut-off on the density of states.  We will show in section \ref{section2} how to use $y^{(p)}(z)$ to derive the $p$-regularized resolvents analogous to~\eqref{eq:regvol}:
\begin{align}
  R_g^{(p)}(-z_1^2,\dots,-z_n^2) = \sum_{k=0}^{2g}\log(p)^{k}r_{g,k}(-z^2_1,\dots,-z^2_n)+\order{p^{-1}}.
  \label{regres}
\end{align}
The loop equations then allow the resolvents to be computed recursively, given the input of the spectral curve \cite{Eynard2018, Eynard2004a, Stanford2019}. Once the regularized resolvents are known,~\eqref{eq:volume} can be used to compute the regularized volumes.

The SFF is then found by taking $\beta_1 = \beta +i t$ and $\beta_2 = \beta -i t$ in the topological expansion of $\ev{Z(\beta_1)Z(\beta_2)}_c$. To compare this expression to the result from universal RMT we will work in the $\tau$-scaled limit defined by taking, $t \to \infty$, $e^{S_0} \to \infty$, but keeping $\tau \coloneqq te^{-S_0} $ finite \cite{Saad2022, Weber2022, Blommaert2022,Okuyama2021, Okuyama2023, Weber2024}. The $\tau$-scaled SFF is then defined as
\begin{align}
\label{1001}
  \kappa(\tau,\beta) \coloneqq \lim_{t \to \infty}\sum_{g=0,\frac{1}{2},1 \dots}^{\infty}\kappa_g(t,\beta)\tau^{2g+1} ,
\end{align}
where:
\begin{align}
  \kappa_g(t,\beta) \coloneqq \frac{Z^{(p)}_g(\beta+it,\beta-it)}{t^{2g+1}}.
\end{align}
In section \ref{section3} we will compute~\eqref{1001} up to $g=1$ and show that it is finite, independent of regularization, and agrees with the prediction of universal RMT.

\subsection{Summary of main results}
\label{summary}
Having established the necessary background, in this section we will summarize the main results of the paper.
\paragraph{Regularized resolvents}  
As explained above, unorientable JT gravity can be regularized by first introducing the $(2,2p+1)$ minimal string model. In section \ref{section2}, we compute the crosscap resolvent of this model as:
\begin{equation}
    \begin{aligned}
        R^{(p)}_{\frac{1}{2}}(z) = \frac{-1}{4z^2} - \sum_{k=1}^{p}\frac{1}{2z\left(z+\frac{(2p+1)}{2\pi}\sin(\frac{\pi k}{2p+1}))\right)}. 
    \end{aligned}
\end{equation}
We then take the asymptotic expansion in $p$ to find the regularized crosscap resolvent of unorientable JT gravity,
\begin{align}
  R^{(p)}_{\frac{1}{2}}(z) = \frac{-1}{4z^2}+\frac{1}{z}H_{2z}-\frac{1}{z}\log(p)+\order{p^{-1}}.
\end{align}
where $H_{2z}$ is the harmonic number. We go on to show how to use the crosscap resolvent to compute higher order resolvents, such as $R^{(p)}_1(z)$ and $R^{(p)}_\frac{1}{2}(z_1,z_2)$. If desired, higher order resolvents can be computed using the same techniques. An important observation we make is that to compute the unorientable volumes of a given $(g,n)$, the entire resolvent at the same $(g,n)$ is actually unnecessary. For the case of $(g,n)=(1,2)$, we take advantage of this and only compute the necessary contributions of the resolvent.
\paragraph{Unorientable volumes}
Starting from the relationship between the resolvents and the volumes given by \cref{eq:volume}, we derive streamlined formulas to compute the unorientable volumes for one and two boundaries. Referring to the coefficients of \cref{eq:regvol}, the formula for one boundary $(n=1)$ is,
\begin{equation}
      v_{g,k}(b) =\sum_{k_1=0}^{\infty}\underset{z=-\frac{k_1}{2}}{\Res}\frac{ 2zf_{g,k}(z) e^{bz}}{2by(z)}
   + \order{b^{-1}},
   \label{n=1}
\end{equation}
and for two boundaries $(n=2)$,
\begin{equation}
\begin{aligned}
   &v_{g,k}(b_1,b_2) = \\[8pt]&\left(\sum_{k_1=1}^{\infty}\underset{z_2=\frac{k_1}{2}}{\Res}\hspace{.1cm}\underset{z_1=-\frac{k_1}{2}}{\Res}+\sum_{k_2=0}^{\infty}\sum_{k_1=0}^{\infty}\underset{z_2=-\frac{k_2}{2}}{\Res}\underset{z_1=-\frac{k_1}{2}}{\Res}\right) -\frac{4z_1z_2f_{g,k}(z_1,z_2)e^{b_1 z_2}e^{b_2 z_2}}{2b_1 b_2y(z_1)} \\[8pt]
   &+\order{b_1^{-1}}+\order{b_2^{-1}},
   \label{n=2}
\end{aligned}
\end{equation}
where the $f_{g,k}$ are found from the loop equations, see \cref{eq:F}. The $\order{b_i^{-1}}$ indicate that the correct volumes are found by dropping all terms of this order, which is further explained in section \ref{section2}. We note that these formulas reduce the volume computations to computing residues, somewhat analogous to topological recursion for the orientable case, except there are now infinitely many residues. We use~\eqref{n=1} and~\eqref{n=2} to compute the unorientable volumes up to $g=1$ for all values of $k \in (0,2g)$. For example, we find the result for $v_{1,0}(b)$ to be surprisingly simple,
\begin{align}
  v_{1,0}(b)
   &=\frac{7b^2}{48} +\frac{\pi^2}{4}+b\log\left(1+e^{-\frac{b}{2}}\right)
  + 2\log\left(1+e^{-\frac{b}{2}}\right)^2,
  \label{118}
 \end{align}
which was numerically computed in \cite{Stanford2023}. At $n=2$, the results for $k=1$ and $k=2$ are still somewhat simple,
\begin{equation}
    \begin{alignedat}{2}
    &v_{1,2}(b_1,b_2) &&= 2b_1^2+2b_2^2+8\pi^2, \\[8pt]
   & v_{1,1}(b_1,b_2)&&=\frac{b_1^3}{2}+\frac{b_2^3}{2}+\frac{b_1^2b_2}{2}+\frac{b_2^2b_1}{2} +\frac{8\pi^2b_1}{3}+\frac{8\pi^2b_2}{3}+56\zeta(3) \\[8pt] 
   &{} &&-32\text{Li}_3\left(-e^{-\frac{b_2}{2}}\right)-32\text{Li}_3\left(-e^{-\frac{b_1}{2}}\right)+32\text{Li}_3\left(e^{-\frac{b_1}{2}-\frac{b_2}{2}}\right)+16\text{Li}_3\left(e^{\frac{1}{2} \left(b_2-b_1\right)}\right)\\[8pt] 
    &{} &&+16\text{Li}_3\left(e^{\frac{1}{2} \left(b_1-b_2\right)}\right)+4\left(b_1-b_2\right)\left( \text{Li}_2\left(e^{\frac{1}{2} \left(b_2-b_1\right)}\right)-\text{Li}_2\left(e^{\frac{1}{2} \left(b_1-b_2\right)}\right)\right)\\[8pt]
    &{} &&+8\left(b_1+b_2\right) \text{Li}_2\left(e^{-\frac{b_1}{2}-\frac{b_2}{2}}\right),
    \label{119}
\end{alignedat}
\end{equation}
where $\text{Li}_{s}(z)$ is the polylogarithm. We also compute $v_{1,0}(b_1,b_2)$, but the result is very long, see~\eqref{v(1,2,0)}, though it importantly exemplifies the multiple polylogarithmic structure of the unorientable volumes, discussed below.

Mirzakhani showed \cite{Mirzakhani2007} that the \textit{orientable} volumes have the following structure:
\begin{align}
    V_g(b_1,\dots,b_n) = \underset{|\alpha| \leq 3g-3+n}{\sum_{\alpha}}C_{\alpha} \pi^{6g-6+2n-2|\alpha|}b_1^{2\alpha_1}\dots b_n^{2\alpha_n}
    \label{orientable}
\end{align}
where $\alpha=(\alpha_1,\dots,\alpha_n)$, $C_{\alpha}\in\mathbb{Q}_+$, $|\alpha| = \sum_i^{n}\alpha_i$ and the dependence of the coefficient on $g$ is suppressed for the sake of readability. For the unorientable case it will be necessary to define the multiple polylogarithm \cite{Waldschmidt2002},
\begin{align}
    \text{Li}_{s_1,\dots,s_d}(z_1,\dots,z_d) = \sum_{k_1>k_2>\dots>k_d>0}\frac{z_1^{k_1}\dots z_d^{k_d}}{k_1^{s_1}\dots k_d^{s_d}}
    \label{polylog}
\end{align}
and the multiple zeta values \cite{Gil2023, Zhao2016},
\begin{align}
    \zeta(s_1,\dots,s_d) = \sum_{k_1>k_2>\dots>k_d>0}\frac{1}{k_1^{s_1}\dots k_d^{s_d}}.
    \label{zeta}
\end{align}
The weight of~\eqref{polylog} and~\eqref{zeta} is defined as $|s|=\sum_{i=1}^d s_i$, where $d$ is called the depth. We find that the unorientable volumes have a symmetric polynomial part that directly generalizes the structure of the orientable volumes~\eqref{orientable}. Justified by our computations, we can make the following conjecture:
\begin{align}
    P[v_{g,k}(b_1,\dots,b_n)] =\underset{|\alpha|+|s|=6g-6+2n-k}{\sum_{\alpha,s}}C_{\alpha,s} \zeta(s_1,\dots,s_d)b_1^{\alpha_1}\dots b_n^{\alpha_n},
    \label{structure}
\end{align}
where $P[v_{g,k}]$ denotes the polynomial part of $v_{g,k}$, $s=(s_1,\dots, s_d)$ with $s_i \in\mathbb{Z}_+$, and the value of $d$ can change in the sum. Here we allow $C_{\alpha,s}$ to be zero, but assume it is positive, i.e. $C_{\alpha,s} \geq 0$, and rational. We also assume $C_{\alpha,s}=0$ when the zeta functions are divergent, e.g. $\zeta(1)$. When $|s|=0$, we take the multi zeta value to lie in $\mathbb{Q}_+$. This formula is further discussed in section \ref{2.6}. As examples, we give the polynomial parts of $v_{1,1}(b_1,b_2)$ and $v_{1,0}(b_1,b_2)$, found in section \ref{section2d},
\begin{equation}
\begin{alignedat}{2}
    &P[v_{1,1}(b_1,b_2)] &&=\frac{b_1^3}{2}+\frac{b_2^3}{2}+\frac{b_1^2b_2}{2}+\frac{b_2^2b_1}{2} +\frac{8\pi^2b_1}{3}+\frac{8\pi^2b_2}{3}+56\zeta(3)\\[8pt]
   &P[v_{1,0}(b_1,b_2)] &&= \frac{5 b_1^4}{96}+\frac{1}{24} b_2 b_1^3+\frac{7}{48} b_2^2 b_1^2+\frac{1}{24} b_2^3 b_1+\frac{5 b_2^4}{96}
   \\[8pt] 
   & &&+\frac{2}{3} \pi ^2\left( b_1^2+b_2^2+\frac{b_1b_2}{2}\right)+4\left( b_1 + b_2 \right)\zeta (3)+\frac{3 \pi ^4}{2}.
\end{alignedat}
\end{equation}
At $|s| \leq 4$, all multi zeta values can be reduced to single zeta values \cite{Gil2023}, which is why there are no multi zeta values in our results. To see multi zeta values, one would have go to at least $|s|=5$. The structure of the polynomial part~\eqref{structure} is a natural generalization of the orientable case since $\pi^{2n} \propto \zeta(2n)$. We note that in the unorientable volumes there are now zeta values of odd integers which is a consequence of the unorientability of the surfaces.\footnote{The connection between zeta values of even and odd integers, orientability, and the rationality of moduli space volumes was discussed in \cite{Witten1991a}.}

More generally, every term in the unorientable volume can be written as a product of a (multiple) polylogarithm and a polynomial. Specifically, each term in $v_{g,k}(b_1,\dots, b_n)$ will have the form,
\begin{align}
   C_{\alpha,s,z} \text{Li}_{s_1,\dots,s_d}(z_1,
    \dots,z_d)b_1^{\alpha_1}\dots b_n^{\alpha_n}, \hspace{.5cm}|s| +|\alpha|=6g-6+2n-k
    \label{totalstructure}
\end{align}
where $C_{\alpha,s,z} \in \mathbb{Q}$ and $z=(z_1,\dots,z_d)$ are exponential functions of $(b_1,\dots,b_n)$ or $\pm 1$\footnote{In all computations we have not seen the $-1$ show up, but it remains a possibility.}. For an illustrative example of this multiple polylogarithmic structure, see~\eqref{v(1,2,0)}, though the volumes in~\eqref{119} and~\eqref{118} also serve as an example\footnote{In~\eqref{118} the $\log(1+e^{-\frac{b}{2}})^2$ term can be written as a linear combination of multi polylogs using a stuffle identity \cite{Waldschmidt2002}.}. We note that the terms in the polynomial part~\eqref{structure} will be a special case of~\eqref{totalstructure}.

\paragraph{SFF from unorientable JT gravity}
The main objective of our work is to compute the late time limit of the SFF by working in the $\tau$-scaled limit, introduced in \cref{background}. The result computed from unorientable JT gravity, using the volumes computed in section \ref{section2}, up to $g=1$, is derived in section \ref{section3a}:
\begin{equation}
    \begin{aligned}
        \kappa(\tau,\beta) & = \kappa_{\text{Airy}}(\tau,\beta)-\frac{2\tau^2}{\sqrt{2\pi \beta}} \sum_{k=1}^{\infty} (-1)^k \left(1-k\sqrt{\frac{\beta \pi}{2}} e^{\frac{\beta k^2}{2}} \text{erfc}\left(k\sqrt{\frac{\beta}{2}}\right)\right) \\[8pt]
  &+\frac{4\tau^3}{\pi}\sum_{k=1}^{\infty}\left(-1+\frac{1}{2}\left(1+\beta k^2\right)e^{\frac{\beta k^2}{2}}E_1\left(\frac{\beta k^2}{2}\right)\right)+\order{\tau^4},
  \label{SFF-jt_intro}
    \end{aligned}
\end{equation}
where the Airy\footnote{Airy corresponds to the low energy limit of JT gravity in the sense that $y^{\text{Airy}}(z) \propto z$.} contribution is 
\begin{align}
\label{airy2}
  \kappa_{\text{Airy}}(\tau,\beta) = \frac{\tau}{2\pi \beta}-\frac{\tau^2}{\sqrt{2\pi \beta}} +\frac{\tau^3}{\pi}\left(\frac{-10}{3}+\log(\frac{2t}{\beta})\right) + \order{\tau^4}.
\end{align}
We see that there are an additional infinite number of terms that arise from considering unorientable JT gravity compared to the Airy model. Immediately, it can be observed that $\kappa(\tau,\beta)$, at least up to $g=1$, is finite, i.e. independent of $\log(p)$. The fact that this is true stems from non trivial cancellations amongst the divergent parts of the volumes. For example, both $v_{1,2}(b_1,b_2)$ and $ v_{1,1}(b_1,b_2)$ contain terms that could in principle contribute to $\kappa(\tau,\beta)$, however, the contributing terms exactly cancel. These cancellations imply that the divergent parts of $V^{(p)}_{g}(b_1,b_2)$ do not contribute to $\kappa(\tau,\beta)$. Furthermore, with the exception of the $\log(\frac{2t}{\beta})$ in $\kappa_\text{Airy}(\tau,\beta)$, we find no terms that break the $\tau$-scaling, i.e. are $t-$dependent. This involves cancellations of terms that would break $\tau-$scaling akin to what was shown in the orientable case \cite{Weber2022, Blommaert2022}, and suggest further constraints on the structure of the unorientable volumes. The problem of $t-$dependence in the $\tau-$scaled limit of the unorientable Airy model was resolved in \cite{Weber2024}.

\paragraph{SFF from universal RMT}
The $\tau-$scaled SFF from universal RMT is given by the Laplace transform of the \textit{microcanonical} SFF:
\begin{align}
 \kappa^{\text{GOE}}(\tau,\beta) = \int_{0}^{\infty}\dd E e^{-2\beta E} \kappa_E^{\text{GOE}}(\tau),
 \end{align}
 with 
 \begin{align}
 \kappa_E^{\text{GOE}}(\tau) = \rho_0(E)\left(1-b^{\text{GOE}}\left(\frac{\tau}{2\pi \rho_0(E)}\right)\right)
\end{align}
where $b^{\text{GOE}}(x)$ is the form factor which can be found in \cite{Mehta2004}. The result is universal in the sense that the functional form of $b^{\text{GOE}}(x)$ only depends on the symmetry class. However, the result will depend on the leading order density of states, $\rho_0(E)$. The density of states for JT gravity is $\rho_0(E) \propto \sinh(2\pi \sqrt{E})$\footnote{In our prior work \cite{Weber2024} the integral was solved to all orders in $\tau$ for the Airy model with $\rho_0(E) \propto \sqrt{E}.$}. The symmetry class, which determines the form factor, is taken to be orthogonal, i.e. GOE. In section \ref{section3b}, we show how to solve this integral order by order in $\tau$. We find the solution up to $\tau^3$ to be: 
\begin{equation}
\begin{aligned}
\kappa^{\text{GOE}}(\tau,\beta)&= \kappa_{\text{Airy}}^{\text{GOE}}(\tau,\beta)-\frac{2\tau^2}{\sqrt{2\pi \beta}} \sum_{k=1}^{\infty} (-1)^k \left(1-k\sqrt{\frac{\beta \pi}{2}} e^{\frac{\beta k^2}{2}} \text{erfc}\left(k\sqrt{\frac{\beta}{2}}\right)\right) \\[8pt] 
& +\frac{4 \tau^3}{\pi}\sum_{k=1}^{\infty}
   \left(-1+\frac{1}{2}e^{\frac{\beta k^2}{2}}(1+\beta k^2)E_1\left(\frac{\beta k^2}{2}\right)\right)+\order{\tau^4},
   \label{GOE_intro}
\end{aligned} 
\end{equation}
with the Airy contribution given by:
\begin{align}
\label{airy1}
  \kappa_{\text{Airy}}^{\text{GOE}}(\tau,\beta) = \frac{\tau}{2\pi \beta}-\frac{\tau^2}{\sqrt{2\pi \beta}}+\frac{-\tau^3}{\pi}\left(\gamma+\log(2\beta \tau^2) + \frac{1}{3}\right) + \order{\tau^4}.
\end{align}
We would like to emphasize the agreement between the second and third terms in \eqref{GOE_intro} and \eqref{SFF-jt_intro}, as an indicator of quantum chaos. In some sense, the agreement is to be expected since JT gravity is dual to a matrix model. However, the two computations are entirely different, as one involves the computations of the complicated unorientable volumes, while the other is just the Laplace transform of the GOE microcanonical SFF. Furthermore, the agreement we observe was only possible due to the cancellations within the divergent parts of the unorientable volumes. Consequently, the presence of this type of cancellations is a necessary condition for the presence of quantum chaos in unorientable JT gravity. Or put differently, one can say that the constraints are an implication of universality in quantum chaotic systems. Note that comparing the Airy terms,~\eqref{airy1} and~\eqref{airy2}, there appears to be disagreement in the $\tau^3$ coefficient. In our prior work \cite{Weber2024}, by computing the Airy terms up to $g=7/2$ and using asymptotic expansions of generalized hypergeometric functions, we show the two series can be made to agree. 

\section{Loop equations and the unorientable volumes}
\label{section2}
The aim of this section is to compute the unorientable volumes from the loop equations. We first introduce how to regularize the resolvents using the $(2,2p+1)$ minimal string model. We show that computing the unorientable volumes from the loop equations reduces to computing an infinity of residues, and introduce a streamlining procedure to facilitate such computations. We compute the unorientable volumes, directly from the regularized resolvents, for the cases $(g,n)=\{(\frac{1}{2},1),(1,1),(\frac{1}{2},2),(1,2)\}$, and show the general structure involves multiple polylogarithms and zeta values of even and odd integers.
\subsection{Preliminaries}
For an introduction and detailed treatment of the loop equations see \cite{Eynard2018, Eynard2016, Eynard2004a, Stanford2019}. 
We will follow the notation of \cite{Stanford2019}, where it was shown the resolvents of a matrix model with orthogonal symmetry satisfy the following recursion\footnote{Here we use the double cover coordinate: $x_i = -z_i^2$, see \cite{Weber2024} for a derivation of the required contour deformation.}:
\begin{align}
\begin{aligned}\label{eq:recursion}
  R_g(-z^2,I)=\frac{1}{2\pi i z}\int_{i\mathbb{R}+\epsilon}\frac{{z'}^2\dd z'}{{z'}^2-z^2}\frac{1}{y(z')}F_g(-z'^2,I),
\end{aligned}
\end{align}
with,
\begin{align}\label{eq:F}
  \begin{aligned}
			F_g(-z^2,I)=&\frac{1}{2z}\partial_z R_{g-\frac 12}(-z^2,I)+R_{g-1}(-z^2,-z^2,I)\\
					&+\sum'_{I\supseteq J, h} R_h(-z^2,J)R_{g-h}(-z^2,I\backslash J)\\
					&+2\sum_{k=1}^{n}\qty[R_0(-z^2,-z^2_k)+\frac{1}{\qty(z_k^2-z^2)^2}]R_{g}(-z^2,I\backslash\qty{-z^2_k}),
			\end{aligned}
\end{align}
where $\sum'$ indicates that $R_0(-z^2)$ and $R_0(-z^2,-z_k^2)$ are excluded from the sum, and 
\begin{align}
  I = \{-z_1^2,\dots,-z_n^2\}.
\end{align}
 The recursion can be solved once given the spectral curve, $y(z)$, along with three special cases that are true for any double scaled one-cut matrix model:
\begin{align}
  R_0(-z_1^2,-z_2^2) = \frac{1}{2z_1 z_2 (z_1+z_2)^2}, \hspace{.5cm} R_0(-z_1^2,-z_2^2,-z_3^2) = \frac{-1}{2z_1^3 z_2^3z_3^3},\hspace{.5cm} R_{0}(-z^2) \sim y(z).
  \label{special}
\end{align}
Note the last case is subtle in the double cover coordinate, since the equality is up to terms analytic around the branch cut in the $x$-coordinate system, with $x=-z^2$. Nevertheless, this property allows the replacement of $R_0(-z^2)$ with $y(z)$ in~\eqref{eq:F}, see \cite{Stanford2019}. From now on, we will make the dependence of $z$ implicit and simply write $R_g(z,I), F_g(z,I)$, etc. We are interested in solving the recursion given by the unorientable JT gravity spectral curve~\eqref{eq:sinh}.

\subsection{The crosscap resolvent $R^{(p)}_{\frac{1}{2}}(z)$}
The divergence of the unorientable volumes is caused by the contribution of small crosscaps, specifically the crosscap is associated with the $(g,n) = (\frac{1}{2},1)$ unorientable volume and has the following measure \cite{Stanford2023}:
\begin{align}
  V_{\frac{1}{2}}(b)bdb = \frac{1}{2}\coth(\frac{b}{4})db.
  \label{coth}
\end{align}
The divergence of the crosscap measure as $b \to 0$, leads to further divergences when this term is iterated in Stanford's recursion. To cure this problem on the loop equations side, the crosscap resolvent $R_{\frac{1}{2}}(z)$ needs to be regularized. As stated in section \ref{background}, this can be done by introducing the spectral curve of the $(2,2p+1)$ minimal string model given by~\eqref{eq:yp} and then taking the asymptotic expansion in $p$. Using~\eqref{eq:recursion} for the case $(g,n) = (\frac{1}{2},1)$ with~\eqref{eq:yp}:
\begin{align}
   R^{(p)}_\frac{1}{2}(z)=\frac{1}{4\pi i z}\int_{i\mathbb{R}+\epsilon}\frac{{z'}\dd z'}{{z'}^2-z^2}\frac{\partial_{z'} y^{(p)}(z')}{y^{(p)}(z')},
   \label{eq:r12}
\end{align}
where $R_0(z)$ was replaced with $y^{(p)}(z)$\footnote{This integral has been written down before, c.f. \cite{Stanford2019, Mertens2021}, though to our knowledge it has not been solved.}. The integrand vanishes at infinity so the integral can be solved by closing a contour on the right half of the plane, orientated clockwise, and computing the residues. Note that the zeroes of $y^{(p)}(z)$ are:
\begin{align}
  y^{(p)}(z_0)= (-1)^p\frac{1}{4\pi}T_{2p+1}\left(\frac{2\pi z_0}{2p+1}\right)=0 \rightarrow z_0 = \frac{2p+1}{2\pi}x_k,
\end{align}
where $x_k$ are the zeroes of the Chebychev polynomial of the first kind:
\begin{align}
  x_k = \sin(\frac{\pi k}{2p+1}) ,\hspace{.5cm} k = -p,\dots,0,\dots,p.
\end{align}
It is useful to note the following expansion:
\begin{align}
  \frac{\partial_{z} y^{(p)}(z)}{y^{(p)}(z)}= \frac{1}{z}+\sum_{k=1}^{p}\frac{2z}{(z-\frac{2p+1}{2\pi}x_k)(z+\frac{2p+1}{2\pi}x_k)}.
  \label{eq:expansion}
\end{align}
Plugging~\eqref{eq:expansion} into~\eqref{eq:r12} and computing the residues in the right half plane\footnote{Here and in the following computations we use the assumption $\forall_i \Re(z_i)>0$ that is implicit in the computation of the resolvents using the loop equations \cite{Weber2024} and amounts to a choice of branch. Note however that after the computation of the resolvents the range of the $z_i$ can be extended to the whole of $\mathbb{C}$ again.}, taking care to include the negative sign for the clockwise orientation of the contour:

\begin{equation}
    \begin{aligned}
        R^{(p)}_{\frac{1}{2}}(z) &= \frac{1}{4\pi i z}\int_{i\mathbb{R}+\epsilon}\frac{{z'}\dd z'}{{z'}^2-z^2}\left(\frac{1}{z'}+\sum_{k=1}^{p}\frac{2z'}{(z'-\frac{(2p+1)}{2\pi}x_k)(z'+\frac{(2p+1)}{2\pi}x_k)}\right) \\
   &= \frac{-1}{4z^2} - \sum_{k=1}^{p}\left(\frac{z}{2z(z-\frac{(2p+1)}{2\pi}x_k)(z+\frac{(2p+1)}{2\pi}x_k)} + \frac{\frac{(2p+1)}{2\pi}x_k}{2z\left((\frac{2p+1}{2\pi}x_k)^2-z^2\right)} \right)\\
   &= \frac{-1}{4z^2} - \sum_{k=1}^{p}\frac{1}{2z(z+\frac{(2p+1)}{2\pi}x_k)}. 
   \label{210}
    \end{aligned}
\end{equation}
Note, that up to now the result is valid for all values of $p$. To find the $p$-regularized resolvent in the form of~\eqref{regres} the asymptotic expansion in $p$ needs to be taken. First, noting that:
\begin{align}
\frac{(2p+1)}{2\pi}x_k = \frac{k}{2} +\order{p^{-1}},
\end{align}
results in:

\begin{equation}
    \begin{aligned}
        R^{(p)}_{\frac{1}{2}}(z)&=  \frac{-1}{4z^2} - \sum_{k=1}^{p}\frac{1}{2z(z+\frac{k}{2})} +\order{p^{-1}}\\[7pt]
  & = \frac{-1}{4z^2} -\frac{\psi ^{(0)}(1+2 z+p)-\psi ^{(0)}(1+2 z)}{z} +\order{p^{-1}}\\[7pt]
  &=\frac{-1}{4z^2} +\frac{1}{z}\left(H_{2z}-\gamma\right)-\frac{1}{z}\log(p) +\order{p^{-1}}.
    \end{aligned}
\end{equation}
To get the last line we used:
\begin{align}
 &\psi^{(0)}(1+2z+p) = \log(p) +\order{p^{-1}} ,\\[7pt]
 & \psi^{(0)}(1+2z)=H_{2z}-\gamma,
\end{align}
where $\psi^{(0)}(x)$ is the digamma function, $H_x$ is the harmonic number, and $\gamma$ is the Euler-Mascheroni constant. It is convenient to make the redefinition: $pe^{\gamma} \to p$, which leads to:  
\begin{align}
  R^{(p)}_{\frac{1}{2}}(z) = \frac{-1}{4z^2}+\frac{1}{z}H_{2z}-\frac{1}{z}\log(p)+\order{p^{-1}}.
  \label{eq:crosscapr}
\end{align}
Analogous to the situation in Stanford's recursion, further divergences in the loop equations will only occur from iterating the crosscap resolvent. Therefore, the loop equations can now be solved with the JT gravity spectral curve~\eqref{eq:sinh}, as long as the regularized crosscap resolvent is used. 

The corresponding regularized unorientable volume can now be computed using~\eqref{eq:volume}:
\begin{equation}
    \begin{aligned}
        bV^{(p)}_{\frac{1}{2}}(b) 
  &=\mathcal{L}^{-1}\qty[\qty( -2z)R^{(p)}_{\frac{1}{2}}(z),\qty(b)] \\ 
&=(-1)\int_{\delta +i\mathbb{R}} \frac{dz}{2\pi i}R^{(p)}_{\frac{1}{2}}(z)(2z)e^{b z} =\\
  &=\frac{1}{2}+2\log(p)\delta(b) -\sum_{k=1}^{\infty}\left(\frac{2}{k}\delta(b)-e^{-\frac{kb}{2}}\right) +\order{p^{-1}}.
  \label{crosscap}
    \end{aligned}
\end{equation}
The delta functions arise from the inverse Laplace transforms of constants. The regularization of $ V^{(p)}_{\frac{1}{2}}(b)$ in~\eqref{crosscap} differs in appearance from the theta function regularization used in \cite{Stanford2023}, but they produce consistent results. For example, the volume $V_\frac{1}{2}^{(p)}(b_1,b_2)$ is a special case in Stanford's recursion and can be computed by integrating over the size of a crosscap \cite{Saad2022, Stanford2023}:
\begin{align}
\label{v121}
 V_\frac{1}{2}^{(p)}(b_1,b_2)= 2\int_0^{a}dbb V_{\frac{1}{2}}^{(p)}(b),
\end{align}
with the condition:
\begin{align}
  2\sinh^2\left(\frac{a}{4}\right)=\cosh(\frac{b_1}{2})+\cosh(\frac{b_2}{2}).
\end{align}
The solution of~\eqref{v121} using~\eqref{crosscap} is:
\begin{align}
  \label{v123}
  V_\frac{1}{2}^{(p)}(b_1,b_2) = 2\log(\frac{\cosh(\frac{b_1}{2})+\cosh(\frac{b_2}{2})}{2})+4\log(2p) +\order{p^{-1}}
\end{align}
which agrees with the results in \cite{Saad2022,Stanford2023} using the theta function regularization with the identification $p \to \frac{2}{\epsilon}$. In section \ref{section2c} this volume will be computed again using the loop equations and~\eqref{v123} will serve as a check on the result.

\subsection{Computation of $V^{(p)}_1(b)$}
\label{sec3}
The regularized volume for the case of $(g,n) = (1,1)$ , i.e. $V^{(p)}_1(z)$, can now be computed. This is the first non-trivial case where we will introduce the streamlining procedure. The starting point is again~\eqref{eq:recursion}:
\begin{align}
   R^{(p)}_1(z)=\frac{1}{2\pi i z}\int_{i\mathbb{R}+\epsilon}\frac{{z'}^2\dd z'}{{z'}^2-z^2}\frac{1}{y(z')}F_1^{(p)}(z').
   \label{eq:R11}
\end{align}
Notice now we are using the JT gravity spectral curve
\begin{align}
  y(z) = \frac{\sin(2\pi z)}{4\pi}.
\end{align}
The notation $F_1^{(p)}(z)$ signifies we have to input the $p$-regularized crosscap resolvent~\eqref{eq:crosscapr} into the definition~\eqref{eq:F}:
\begin{align}
 F_1^{(p)}(z) = \frac{1}{2z}\partial_{z}R^{(p)}_{\frac{1}{2}}(z) + R_{0}(z,z)+R^{(p)}_{\frac{1}{2}}(z)^2 ,
 \label{eq:F11}
\end{align}
where $R_0(z,z)$ is given from the special cases~\eqref{special}. To compute the integral~\eqref{eq:R11} the contour is closed in the right half plane. Since $R^{(p)}_{\frac{1}{2}}(z)$ only has poles on the negative real axis and zero, as seen from~\eqref{eq:crosscapr},
the only residues that need to be computed are from $z'=z$ and the residues from the spectral curve. The inverse spectral curve has the following expansion:
\begin{align}
   \frac{1}{y(z)} = \frac{2}{z}+\sum_{k=1}^{\infty}\frac{4z(-1)^k}{(z-\frac{k}{2})(z+\frac{k}{2})},\label{eq:expansion2}
\end{align}
so the residues at the positive half integers need to be computed.
These considerations lead to the following solution of the integral:
\begin{equation}
    \begin{aligned}
        R^{(p)}_1(z)&=\frac{1}{2\pi i z}\int_{i\mathbb{R}+\epsilon}\frac{{z'}^2\dd z'}{{z'}^2-z^2}\frac{1}{y(z')}F_1^{(p)}(z') \\[10pt]
&= -\frac{F_1^{(p)}(z)}{2y(z)}+\sum_{k=1}^{\infty}\frac{2(-1)^k\left(\frac{k}{2}\right)^2 F_1^{(p)}(\frac{k}{2})}{z(z-\frac{k}{2})(z+\frac{k}{2})}.
\label{eq:R11_F}
    \end{aligned}
\end{equation}
The first term is simply the residue at $z'=z$ and the second term comes from the residues at the positive half integers. The corresponding volume is then given by
\begin{align}
 V_1^{(p)}(b)=(-1)\int_{\delta +i\mathbb{R}} \frac{dz}{2\pi i}\frac{2z R^{(p)}_{1}(z)e^{b z}}{b}.
 \label{eq:v11_1}
\end{align}
By expanding $y(z)^{-1}$ using~\eqref{eq:expansion2} it is seen $R_1^{(p)}(z)$ only has poles on the negative real axis and zero. The inverse Laplace transform~\eqref{eq:v11_1} is given by closing the contour in the negative half plane and computing the residues at $z=0$ and $z=-\frac{k}{2}$, where $k$ is an integer:
\begin{align}
  V_1^{(p)}(b) = -\left(\underset{z=0}{\Res}+\sum_{k=1}^{\infty}\underset{z=-\frac{k}{2}}{\Res}\right)\frac{2z R^{(p)}_{1}(z)e^{b z}}{b}  \, .
  \label{eq:v11_2}
\end{align}
A basic property of the regularized volumes is that, by definition, they cannot be divergent as $b \to 0$. The only source of the divergence, the crosscap volume, has been regularized so no other unorientable volume can diverge in this limit. However, many of the poles in~\eqref{eq:v11_2} are of order one and will produce terms that go as $\order{b^{-1}}$. There are also higher order poles that will produce terms that go as $\order{b^{-1}}$ when the derivative does not act on $e^{bz}$. These terms then have to mutually cancel to assure that $V_1^{(p)}(b)$ is finite as $b \to 0$. To further elaborate on this point, the structure of the residues in~\eqref{eq:v11_2} restricts the form of the $\order{b^{-1}}$ terms to being either $\propto b^{-1} $  or $\propto b^{-1} e^{-{\frac{bk}{2}}}$. No combination of terms such as these can form a term that is purely finite as $b \to 0$, unless they all cancel\footnote{If there were also terms that were $ \propto b^{-1}e^{\frac{bk}{2}}$ it would be possible to construct an identity that does this, for example using polylogarithms \cite{maximon2003}.}.
We will thus not compute any of the $\order{b^{-1}}$ terms. Immediately this allows the replacement
\begin{align}
  R^{(p)}_1(z) \rightarrow -\frac{F_1^{(p)}(z)}{2y(z)},
\end{align}
since the second term in~\eqref{eq:R11_F} only has order one poles. It should be stressed the replacement is only valid if we do not compute any of the $\order{b^{-1}}$ terms. The volume calculation then becomes
\begin{align}
\label{eq:v_11_0}
   V_1^{(p)}(b) &=-\left(\underset{z=0}{\Res}+\sum_{k=1}^{\infty}\underset{z=-\frac{k}{2}}{\Res}\right)\left(\frac{ -2zF^{(p)}_1(z) e^{bz}}{2y(z)b}\right)
   + \order{b^{-1}}.
\end{align}
A detailed solution of~\eqref{eq:v_11_0} is presented in Appendix \ref{details1}, but it be should emphasized the computation is greatly simplified by ignoring all of the $\order{b^{-1}}$ terms. The expression for the total volume is
 \begin{align}
   V_1^{(p)}(b)&= \sum_{k=0}^{2}\log(p)^{k}v_{1,k}(b)+\order{p^{-1}}\\[8pt]
   &=2\log(p)^2 +\left(b+4\log(1+e^{-\frac{b}{2}})\right)\log(p)\\[8pt] \nonumber
   &+\frac{7b^2}{48} +\frac{\pi^2}{4}+b\log\left(1+e^{-\frac{b}{2}}\right)
  + 2\log\left(1+e^{-\frac{b}{2}}\right)^2 +\order{p^{-1}}.
 \end{align}
 The results for $v_{1,2}(b)$ and $v_{1,1}(b)$ were reported in \cite{Stanford2023} and agree with the results here with the identification $p \to \frac{2}{\epsilon}$.\footnote{The expression for $v_{1,1}(b)$ given in \cite{Stanford2023} can be simplified using a polylogarithm identity \cite{maximon2003}.} The value for $v_{1,0}(b)$ was numerically computed in \cite{Stanford2023} and can be shown to be consistent with the analytic result presented here by plotting $v_{1,0}(b)$.\footnote{For the plots to match the additional terms arising from the change $p \to \frac{2}{\epsilon}$ have to be accounted for.}.
 \subsection{Computation of $V^{(p)}_\frac{1}{2}(b_1,b_2)$}
 \label{section2c}
 The computation of $V^{(p)}_\frac{1}{2}(b_1,b_2)$ is actually much simpler than that of $V^{(p)}_1(b)$ and it has already been computed using Stanford's recursion, but it exemplifies an important simplification for the loop equations that occurs for cases with $n>1$. Beginning again with~\eqref{eq:recursion},
\begin{align}
   R^{(p)}_{\frac{1}{2}}(z_1,z_2)=\frac{1}{2\pi i z_1}\int_{i\mathbb{R}+\epsilon}\frac{{z'}^2\dd z'}{{z'}^2-z_1^2}\frac{1}{y(z')}F^{(p)}_\frac{1}{2}(z',z_2),
   \label{R122}
\end{align}
where,
\begin{align}
\label{F1/2}
  F^{(p)}_{\frac{1}{2}}(z_1,z_2) =\frac{1}{2z_1}\partial_{z_1} R_{0}(z_1,z_2) +2R^{(p)}_{\frac{1}{2}}(z_1)\left(R_{0}(z_1,z_2)+\frac{1}{(z_1^2-z_2^2)^2}\right).
\end{align}
Accounting for the fact the resolvents only have poles in the negative half plane, the solution to~\eqref{R122} is
\begin{align}
\label{R122f}
   R^{(p)}_{\frac{1}{2}}(z_1,z_2) = &\frac{-F^{(p)}_{\frac{1}{2}}(z_1,z_2)}{2y(z_1)} +\sum_{k=1}^{\infty}\frac{2(-1)^k\left(\frac{k}{2}\right)^2F^{(p)}_{\frac{1}{2}}(\frac{k}{2},z_2)}{z_1(z_1-\frac{k}{2})(z_1+\frac{k}{2})}+ \\
  \nonumber &\frac{1}{y(z_2)}R^{(p)}_{\frac{1}{2}}(z_2)\left(R_{0}(z_1,z_2)+\frac{1}{(z_1^2-z_2^2)^2}\right)-\frac{1}{2z_1(z_2^2-z_1^2)}\left(\partial_{z_2}\frac{R^{(p)}_{\frac{1}{2}}(z_2)}{y(z_2)}\right).
\end{align}
$R^{(p)}_{\frac{1}{2}}(z_1,z_2)$ is symmetric with respect to $z_1 \leftrightarrow z_2$ and has no poles in the positive half plane, though it is not obvious from this form. The corresponding unorientable volume is
\begin{align}
V_{\frac{1}{2}}(b_1,b_2) = \int \int_{i\mathbb{R}+\delta} dz_1 dz_2 \frac{4z_1z_2 R_{\frac{1}{2}}(z_1,z_2)}{b_1 b_2}e^{b_1z_1}e^{b_2 z_2}.
\label{Vb1b2}
\end{align}
The strategy to compute this expression is to first compute the inverse Laplace transform with respect $z_1$ and then with respect to $z_2$. Based on the discussion of section \ref{sec3}, the second term in~\eqref{R122f} can be dropped since it only contains order one poles of $z_1$. There are also poles at $z_1=-z_2$, some of which will be of order greater than one. However, consider what would happen if we computed a pole at $z_1=-z_2$. After computing subsequent residues of $z_2$, it would necessarily produce a term like
\begin{align}
  \propto \frac{b_1^m (b_2-b_1)^n \theta(b_2-b_1)}{b_2 b_1}.
  \label{example}
\end{align}
The $m$ and $n$ are arbitrary and only signify the order of the poles where the residues were computed. This term does have contributions that are finite as $b_i \to 0$, but it also contains a term of order $\order{b_2^{-1}}$ that must cancel. Due to the presence of the $\theta(b_2-b_1)$ the only terms that could cancel the $\order{b_2^{-1}}$ term are expressions with the exact same structure as~\eqref{example} and therefore the entirety of~\eqref{example} would cancel. Thus, none of the residues at $z_1 = -z_2$ have to be computed. The last line in~\eqref{R122f} can now be dropped, leading to the replacement in~\eqref{Vb1b2}:
\begin{align}
  R^{(p)}_{\frac{1}{2}}(z_1, z_2) \rightarrow \frac{-F^{(p)}_{\frac{1}{2}}(z_1,z_2)}{2y(z_1)}.
\end{align}
The contour for the $z_1$ integral can be closed in the negative half plane and only the residues at $z_1=0$ and $z_1=-\frac{k}{2}$ need to be computed. After taking the residues at $z_1=-\frac{k}{2}$, the result can have poles at $z_2 = \pm \frac{k}{2}$ because of the poles at $z_2 = \pm z_1$\footnote{It is true the resolvent has no poles at $z_2=z_1$, or anywhere on the positive axis, but after computing the residues of $z_1$ this no longer holds.}, and poles at $z_2=0$ but these residues will all be $\order{b_2^{-1}}$ terms. These considerations lead to the following expression for the volume:
\begin{equation}
    \begin{aligned}
        \label{v1/2}
  V^{(p)}_{\frac{1}{2}}(b_1,b_2) &= \left(\underset{z_2=0}{\Res}\hspace{.1cm}\underset{z_1=0}{\Res}+\sum_{k=1}^{\infty}\underset{z_2=\pm \frac{k}{2}}{\Res}\underset{z_1=-\frac{k}{2}}{\Res}\right)\frac{-4z_1z_2 F^{(p)}_{\frac{1}{2}}(z_1,z_2)}{2y(z_1)b_1 b_2}e^{b_1z_1}e^{b_2 z_2} \\
  &+\order{b_1^{-1}}+\order{b_2^{-1}}.
    \end{aligned}
\end{equation}
It should be stressed the residues of $z_1$ and $z_2$ do not commute with each other and $z_1$ has to be computed first. The residues at $z_2=\frac{+k}{2}$ and $z_2=\frac{-k}{2}$ are implicitly summed over here. To be rigorous, the sum in~\eqref{v1/2} is only convergent when $b_1>b_2$ due to the residues at $z_2=\frac{k}{2}$. We can guarantee convergence of the sum by taking advantage of the symmetry of the volume and writing,
\begin{equation}
    \begin{aligned}
        V^{(p)}_{\frac{1}{2}}(b_1,b_2) =  V^{(p)}_{\frac{1}{2}}(b_1,b_2)\theta(b_1-b_2) +V^{(p)}_{\frac{1}{2}}(b_2,b_1)\theta(b_2-b_1.) 
        \label{v1/2_2}
    \end{aligned}
\end{equation}
Using~\eqref{v1/2} and~\eqref{v1/2_2} we get the following expression for the volume where the sum is now convergent:
\begin{equation}
    \begin{aligned}
         V^{(p)}_{\frac{1}{2}}(b_1,b_2) &= \left(\underset{z_2=0}{\Res}\hspace{.1cm}\underset{z_1=0}{\Res}+\sum_{k=1}^{\infty}\underset{z_2=\pm \frac{k}{2}}{\Res}\underset{z_1=-\frac{k}{2}}{\Res}\right)\frac{-4z_1z_2 F^{(p)}_{\frac{1}{2}}(z_1,z_2)}{2y(z_1)b_1 b_2}e^{b_1z_1}e^{b_2 z_2}\theta(b_1-b_2) \\
  & +(b_1 \leftrightarrow b_2)+\order{b_1^{-1}}+\order{b_2^{-1}}
  \label{v1/2_3}
    \end{aligned}
\end{equation}
where we have added the result with $b_1$ and $b_2$ switched in accordance with~\eqref{v1/2_2}.
However, once the sums are computed, the theta functions can be dropped since the volume is symmetric, so this does not present any additional complications\footnote{In practice, one does not have to worry about the theta functions. The sums just have to be computed assuming $b_1>b_2$ and then analytically continued to all values of $(b_1, b_2)$. The final result will be symmetric. }.
The details of the computation are left for Appendix \ref{details2}, but we again emphasize the computation is much simpler once all of the terms of order $\order{b_1^{-1}}$ and $\order{b_2^{-1}}$ are ignored. The final result is
\begin{equation}
    \begin{aligned}
        \label{1}
 V^{(p)}_{\frac{1}{2}}(b_1,b_2) &= 4\log(p) + b_1+ 2\log(1+e^{\frac{b_2-b_1}{2}})+2\log(1+e^{-\frac{b_2+b_1}{2}}) +\order{p^{-1}}\\
 & = 2\log(\frac{\cosh(\frac{b_1}{2})+\cosh(\frac{b_2}{2})}{2})+4\log(2p)+\order{p^{-1}},
    \end{aligned}
\end{equation}
which agrees with the result computed from Stanford's recursion~\eqref{v123}. 

We note that we can also write $v_{\frac{1}{2},0}(b_1,b_2)$ as
\begin{align}
   v_{\frac{1}{2},0}(b_1,b_2)= \frac{b_1}{2}+\frac{b_2}{2}+ \log(1+e^{\frac{b_2-b_1}{2}})+ \log(1+e^{\frac{b_1-b_2}{2}})+2\log(1+e^{-\frac{b_2+b_1}{2}}),
\end{align}
which makes the symmetric polynomial part, introduced in \ref{summary}, obvious:
\begin{align}
    P[v_{\frac{1}{2},0}(b_1,b_2)]=\frac{b_1}{2}+\frac{b_2}{2}.
\end{align}
\subsubsection{An Airy aside}
As an aside we note this demonstrates a simpler way to compute the unorientable Airy volumes that were computed in \cite{Weber2024}. For example, writing $V^{(p)}_{\frac{1}{2}}(b_1,b_2)$ as in~\eqref{v1/2_2},
\begin{equation}
    \begin{aligned}
        V^{(p)}_{\frac{1}{2}}(b_1,b_2) =V^{(p)}_{\frac{1}{2}}(b_1,b_2)\theta(b_1-b_2) + V^{(p)}_{\frac{1}{2}}(b_2,b_1)\theta(b_2-b_1),
  \label{symmetry}
    \end{aligned}
\end{equation}
allows the Airy terms to be read off as the leading order terms from~\eqref{1}:
\begin{align}
  V_{\frac{1}{2}}^{\text{Airy}}(b_1,b_2) = b_1\theta(b_1-b_2)+b_2\theta(b_2-b_1).
\end{align}
More generally this method allows the Airy terms for any genus to be computed by using $F_g(z_1,z_2)$ given by the Airy spectral curve and taking the residues at zero:
\begin{equation}
    \begin{aligned}
        V_g^{\text{Airy}}(b_1,b_2)=\underset{z_2=0}{\Res}\hspace{.1cm}\underset{z_1=0}{\Res}\frac{-4z_1z_2 F^{\text{Airy}}_{g}(z_1,z_2)e^{b_1z_1}e^{b_2 z_2}}{2y^{\text{Airy}}(z_1)b_1 b_2}\theta(b_1-b_2) +\left(b_1 \leftrightarrow b_2\right).
    \end{aligned}
\end{equation}
This formula leads to more efficient implementation since it does not require taking any residues at $z_1=-z_2$. It is likely this formula can be generalized to higher boundaries, but we do not pursue this here.
\subsection{Computation of $V_1^{(p)}(b_1,b_2)$}
\label{section2d}
Based on the discussion of the previous two sections we know all of the information necessary to compute $V_1^{(p)}(b_1,b_2)$ is contained in
\begin{align}
\label{F_1(z1,z2)}
  -\frac{4z_1z_2F^{(p)}_{1}(z_1,z_2)e^{b_1 z_2}e^{b_2 z_2}}{2b_1 b_2y(z_1)},
\end{align}
where, 
\begin{equation}
    \begin{aligned}
        F^{(p)}_1(z_1,z_2)=& \frac{1}{2z_1}\partial_{z_1}R^{(p)}_{\frac{1}{2}}(z_1,z_2) + R_{0}(z_1,z_1,z_2) +\\
  &2R^{(p)}_1(z_1)\left(\frac{1}{2}\frac{1}{z_1 z_2 \qty(z_1+z_2)^2}+\frac{1}{(z_1^2-z_2^2)^2}\right) +2R^{(p)}_{\frac{1}{2}}(z_1)R^{(p)}_{\frac{1}{2}}(z_1,z_2).
    \end{aligned}
\end{equation}
To compute $V_1^{(p)}(b_1,b_2)$ we have to take various combinations of residues at $z_1=0, \frac{-k_1}{2}$ and $z_2=0, \frac{+k_1}{2}, \frac{-k_2}{2}$, with $k_2$ not necessarily equal to $k_1$ (note the positive case does have to be equal), while ignoring all of the terms that go as $\order{b_1^{-1}} $ and $\order{b_2^{-1}}$. We first expand $F^{(p)}_1(z_1,z_2)$ in powers of $\log(p)$,
\begin{equation}
\begin{aligned}
F^{(p)}_1(z_1,z_2)   = \sum_{k=0}^{2}f_{1,k}(z_1,z_2)\log(p)^k,
\end{aligned}
\end{equation}
then the exact formula for the volumes with divergence of order $k$ is:
\begin{equation}
\begin{aligned}
   &v_{1,k}(b_1,b_2) = \\[8pt]&\left(\sum_{k_1=1}^{\infty}\underset{z_2=\frac{k_1}{2}}{\Res}\hspace{.1cm}\underset{z_1=-\frac{k_1}{2}}{\Res}+\sum_{k_2=0}^{\infty}\sum_{k_1=0}^{\infty}\underset{z_2=-\frac{k_2}{2}}{\Res}\underset{z_1=-\frac{k_1}{2}}{\Res}\right) -\frac{4z_1z_2f_{1,k}(z_1,z_2)e^{b_1 z_2}e^{b_2 z_2}}{2b_1 b_2y(z_1)} \\[8pt]
   &+\order{b_1^{-1}}+\order{b_2^{-1}},
\end{aligned}
\end{equation}
where the sum is computed assuming $b_1>b_2$ and then the result is analytically continued to all values of $(b_1,b_2)$ (see discussion surrounding~\eqref{v1/2_3}). Details of the computations are given in Appendix~\eqref{details3} but we will state the results here. \\ The result for $k=2$ is,
\begin{align}
\label{v(1,2,2)}
    v_{1,2}(b_1,b_2) = 2b_1^2+2b_2^2+8\pi^2
\end{align}
for $k=1$,
\begin{equation}
    \begin{aligned}
    \label{v(1,2,1)}
   v_{1,1}(b_1,b_2)&=\frac{1}{3} \left(2 b_1^3+3 b_2^2 b_1+b_2^3+4 \pi ^2 \left(3 b_1+b_2\right)+168 \zeta (3)\right) \\[8pt] 
    &-32\text{Li}_3\left(-e^{-\frac{b_2}{2}}\right)-32\text{Li}_3\left(-e^{-\frac{b_1}{2}}\right)+32\text{Li}_3\left(e^{-\frac{b_1}{2}-\frac{b_2}{2}}\right)+32\text{Li}_3\left(e^{\frac{1}{2} \left(b_2-b_1\right)}\right)\\[8pt] 
    &+8\left(b_1-b_2\right) \text{Li}_2\left(e^{\frac{1}{2} \left(b_2-b_1\right)}\right)+8\left(b_1+b_2\right) \text{Li}_2\left(e^{-\frac{b_1}{2}-\frac{b_2}{2}}\right),
\end{aligned}
\end{equation}
and for k=0,
    \begin{equation}
\begin{aligned}
     v_{1,0}(b_1,b_2) &=\frac{1}{12} b_1 \left(b_2^3+4 \pi ^2 b_2+96 \zeta (3)\right)+\frac{7 b_1^4}{96}+\left(\frac{7 b_2^2}{48}+\frac{5 \pi ^2}{6}\right) b_1^2+\frac{b_2^4}{32}+\frac{b_2^2\pi^2}{2}+\frac{3\pi^4}{2} \\[8pt]
    &-8b_1\text{Li}_{3}\left(-e^{-\frac{1}{2}b_2}\right) -8b_2\text{Li}_{3}\left(e^{-\frac{1}{2}b_1}\right)+\left(2 b_1^2+2b_2^2+2b_1b_2\right) \text{Li}_2\left(e^{-\frac{b_1}{2}-\frac{b_2}{2}}\right) \\[8pt]
    &+8 b_1 \text{Li}_3\left(e^{\frac{1}{2} \left(b_2-b_1\right)}\right)+8\left( b_1 +b_2\right)\left(\text{Li}_3\left(e^{-\frac{b_1}{2}-\frac{b_2}{2}}\right)-\text{Li}_{2,1}\left(e^{-\frac{b_1}{2}-\frac{b_2}{2}},1\right)\right)\\[8pt]
    &+\left(2 b_1^2 -2b_2b_1 \right) \text{Li}_2\left(e^{\frac{1}{2} \left(b_2-b_1\right)}\right)+8 \left(b_2-b_1\right)\text{Li}_{2,1}\left(e^{\frac{1}{2}\left(b_2-b_1\right)},1\right)+16 \text{Li}_4\left(e^{\frac{1}{2} \left(b_2-b_1\right)}\right)\\[8pt]
    &+64\text{Li}_4\left(e^{-\frac{1}{2}\left(b_1+b_2\right)}\right)-32\text{Li}_{3,1}\left(e^{-\frac{1}{2}\left(b_1+b_2\right)},1\right)-32\text{Li}_{3,1}\left(e^{-\frac{1}{2}\left(b_1-b_2\right)},1\right)\\[8pt]
    &+16\text{Li}_{2,2}\left(e^{-\frac{1}{2}\left(b_1+b_2\right)},1\right)+32\text{Li}_{2,2}\left(1,e^{-\frac{1}{2}\left(b_1+b_2\right)}\right)+16\text{Li}_{2,2}\left(1,e^{-\frac{1}{2}\left(b_1-b_2\right)}\right) \\[8pt]
 &  -16\text{Li}_{2,2}\left(-e^{-\frac{1}{2}b_2},e^{-\frac{1}{2}\left(b_1-b_2\right)}\right)+16\text{Li}_{2,2}\left(e^{-\frac{1}{2}\left(b_1+b_2\right)},-e^{\frac{1}{2}b_2}\right) \\[8pt] 
 &+16\text{Li}_{2,2}\left(e^{-\frac{1}{2}\left(b_1+b_2\right)},-e^{-\frac{1}{2}b_2}\right)+16\text{Li}_{2,2}\left(-e^{-\frac{1}{2}b_2},e^{-\frac{1}{2}\left(b_1+b_2\right)}\right)\\[8pt]
 &-16\text{Li}_{3,1}\left(e^{-\frac{1}{2}\left(b_1+b_2\right)},-e^{-\frac{1}{2}b_2}\right) -16\text{Li}_{1,3}\left(-e^{-\frac{1}{2}b_2},e^{-\frac{1}{2}\left(b_1+b_2\right)}\right)\\[8pt]
 &+16\text{Li}_{3,1}\left(e^{-\frac{1}{2}\left(b_1+b_2\right)},-e^{\frac{1}{2}b_2}\right)-16\text{Li}_{1,3}\left(-e^{-\frac{1}{2}b_2},e^{-\frac{1}{2}\left(b_1-b_2\right)}\right)\\[8pt]
 &+32\text{Li}_{3,1}\left(-e^{-\frac{b_1}{2}},-e^{-\frac{b_2}{2}}\right) +32\text{Li}_{1,3}\left(-e^{-\frac{b_2}{2}},-e^{-\frac{b_1}{2}}\right).
 \label{v(1,2,0)}
\end{aligned}
\end{equation}
where $\text{Li}_{s_1,\dots,s_j}(z_1,\dots,z_j)$ is the multi-polylog defined in~\eqref{polylog}. We have numerically verified the symmetry of $v_{1,1}(b_1,b_2)$ and $v_{1,0}(b_1,b_2)$ with respect to $b_1 \leftrightarrow b_2$, which is an important check on the result\footnote{We used PolyLogTools for the numerical verification \cite{Duhr2019}, which uses the opposite convention for multi-polylogs than we do.}. Other checks on these results are that $v_{1,0}(b_1,b_2)$ gives the correct Airy volume in the large $b_1,b_2$ limit, and each volume gives the correct contribution to the SFF discussed in section \ref{section3}.

Since the volumes are necessarily  symmetric we can rewrite them in a way that makes this symmetry manifest. For example, we can make the trivial manipulation,
\begin{equation}
    v_{1,1}(b_1,b_2)=\frac{v_{1,1}(b_1,b_2)}{2}+\frac{v_{1,1}(b_2,b_1)}{2},
    \label{symmetrize}
\end{equation}
to rewrite this volume as:
\begin{equation}
\begin{aligned}
   v_{1,1}(b_1,b_2) &=  \frac{b_1^3}{2}+\frac{b_2^3}{2}+\frac{b_1^2b_2}{2}+\frac{b_2^2b_1}{2} +\frac{8\pi^2b_1}{3}+\frac{8\pi^2b_2}{3}+56\zeta(3) \\[8pt] 
   &-32\text{Li}_3\left(-e^{-\frac{b_2}{2}}\right)-32\text{Li}_3\left(-e^{-\frac{b_1}{2}}\right)+32\text{Li}_3\left(e^{-\frac{b_1}{2}-\frac{b_2}{2}}\right)+16\text{Li}_3\left(e^{\frac{1}{2} \left(b_2-b_1\right)}\right)\\[8pt] 
    &+16\text{Li}_3\left(e^{\frac{1}{2} \left(b_1-b_2\right)}\right)+4\left(b_1-b_2\right)\left( \text{Li}_2\left(e^{\frac{1}{2} \left(b_2-b_1\right)}\right)-\text{Li}_2\left(e^{\frac{1}{2} \left(b_1-b_2\right)}\right)\right)\\[8pt]
    &+8\left(b_1+b_2\right) \text{Li}_2\left(e^{-\frac{b_1}{2}-\frac{b_2}{2}}\right)
    \end{aligned}
\end{equation}
which is the result given in section \ref{summary}. We can then read off the polynomial part to be,
\begin{equation}
     P[v_{1,1}(b_1,b_2)] =  \frac{b_1^3}{2}+\frac{b_2^3}{2}+\frac{b_1^2b_2}{2}+\frac{b_2^2b_1}{2} +\frac{8\pi^2b_1}{3}+\frac{8\pi^2b_2}{3}+56\zeta(3).
     \label{poly}
\end{equation}
We will not write $v_{1,0}(b_1,b_2)$ in a manifestly symmetric form due to its length, but the polynomial part can be read off as:
\begin{equation}
\begin{aligned}
   P[v_{1,0}(b_1,b_2)] &= \frac{5 b_1^4}{96}+\frac{1}{24} b_2 b_1^3+\frac{7}{48} b_2^2 b_1^2+\frac{1}{24} b_2^3 b_1+\frac{5 b_2^4}{96}
   \\[8pt] 
   &+\frac{2}{3} \pi ^2\left( b_1^2+b_2^2+\frac{b_1b_2}{2}\right)+4\left( b_1 + b_2 \right)\zeta (3)+\frac{3 \pi ^4}{2}.
\end{aligned}
\end{equation}
\subsection{General structure of unorientable volumes}
\label{2.6}
Following the discussion of the previous sections, the exact streamlined formulas to compute the unorientable volumes with genus $g$, divergence $k$, and one and two boundaries can now be written down. For one boundary $n=1$ the formula is,
\begin{equation}
      v_{g,k}(b) =\sum_{k_1=0}^{\infty}\underset{z=-\frac{k_1}{2}}{\Res}\frac{ 2zf_{g,k}(z) e^{bz}}{2by(z)}
   + \order{b^{-1}},
   \label{n1}
\end{equation}
and for two boundaries $n=2$,
\begin{equation}
\begin{aligned}
   &v_{g,k}(b_1,b_2) = \\[8pt]&\left(\sum_{k_1=1}^{\infty}\underset{z_2=\frac{k_1}{2}}{\Res}\hspace{.1cm}\underset{z_1=-\frac{k_1}{2}}{\Res}+\sum_{k_2=0}^{\infty}\sum_{k_1=0}^{\infty}\underset{z_2=-\frac{k_2}{2}}{\Res}\underset{z_1=-\frac{k_1}{2}}{\Res}\right) -\frac{4z_1z_2f_{g,k}(z_1,z_2)e^{b_1 z_2}e^{b_2 z_2}}{2b_1 b_2y(z_1)} \\[8pt]
   &+\order{b_1^{-1}}+\order{b_2^{-1}},
   \label{n2}
\end{aligned}
\end{equation}
where again the sum is computed assuming $b_1>b_2$ and then the result is analytically continued to all values of $(b_1,b_2)$. We further emphasize that these formulas give the exact volumes, and the $\order{b_i^{-1}}$ are meant to indicate that the correct volumes are found by ignoring all terms of this order. The $f_{g,k}$ depend on lower order resolvents that can be computed using the loop equations, and will follow analogously to the computation of $R^{(p)}_{\frac{1}{2}}(z_1,z_2)$ done in section \ref{section2c}. It is likely that these formulas can be generalized to arbitrary number of boundaries, but we will not pursue this here.

The polynomial part of the unorientable volumes has a structure that generalizes the structure of the orientable volumes \cite{Mirzakhani2007}. 
We can make precise how to compute the polynomial part for one and two boundaries using~\eqref{n1} and~\eqref{n2}. For one boundary the formula is,
\begin{equation}
    \begin{aligned}
        P[v_{g,k}(b)] =\underset{z=0}{\Res}\frac{ 2zf_{g,k}(z) e^{bz}}{2by(z)}
   + \order{b^{-1}}
   \label{n1a}
    \end{aligned}
\end{equation}
and for two boundaries,
\begin{equation}
\begin{aligned}
   P[v_{g,k}(b_1,b_2)]&=\frac{1}{2}\underset{z_2=0}{\Res}\hspace{.1cm}\underset{z_1=0}{\Res} -\frac{4z_1z_2f_{g,k}(z_1,z_2)e^{b_1 z_2}e^{b_2 z_2}}{2b_1 b_2y(z_1)} +\order{b_1^{-1}}+\order{b_2^{-1}} \\[8pt]
   &+\frac{1}{2}\left(b_1 \leftrightarrow b_2\right).
   \label{n2a}
\end{aligned}
\end{equation}
where we have symmetrized the result, e.g ~\eqref{symmetrize} -~\eqref{poly}. From the computations detailed in prior sections, we can make a conjecture about the structure of the polynomial part of the unorientable volumes, which we generalize to $n$ boundaries:
\begin{align}
    P[v_{g,k}(b_1,\dots,b_n)] =\underset{|\alpha|+|s|=6g-6+2n-k}{\sum_{\alpha,s}}C_{\alpha,s} \zeta(s_1,\dots,s_d)b_1^{\alpha_1}\dots b_n^{\alpha_n},
    \label{structure1}
\end{align}
where $\alpha = (\alpha_1,\dots, \alpha_n)$, $s=(s_1,\dots, s_d)$ with $s_i \in\mathbb{Z}_+$, and the value of $d$ can change in the sum. Here we allow $C_{\alpha,s}$ to be zero, but assume it is positive, i.e. $C_{\alpha,s} \geq 0$, and rational. We also assume $C_{\alpha,s}=0$ when the zeta functions are divergent, e.g. $\zeta(1)$. When $|s|=0$, we take the multi zeta value to lie in $\mathbb{Q}_+$. We note that this definition is a matter of convention. For one and two boundaries we choose to define the polynomial part based on~\eqref{n1a} and~\eqref{n2a}, and we choose it to be symmetric. However, these choices are only inherently necessary insofar that they make the $P[v_{g,k}(b_1,\dots,b_n)]$ a direct generalization of the orientable volumes. 
Furthermore, it is possible the zeta values should be alternating zeta values, however, in all computations done so far the polynomial part is described by~\eqref{structure}.\footnote{Alternating zeta values would allow for $\log(2)$'s, the absence of which is somewhat surprising given the regularization dependence of the volumes.} 

More generally, every term in the unorientable volume can be written as a product of a (multiple) polylogarithm and a polynomial. Specifically, each term in $v_{g,k}(b_1,\dots, b_n)$ will have the form,
\begin{align}
   C_{\alpha,s,z} \text{Li}_{s_1,\dots,s_d}(z_1,
    \dots,z_d)b_1^{\alpha_1}\dots b_n^{\alpha_n},\hspace{.5cm}|s| +|\alpha|=6g-6+2n-k,
    \label{totalstructure1}
\end{align}
where $C_{\alpha,s,z} \in \mathbb{Q}$ and the $z=(z_1,\dots,z_d)$ are exponential functions of $(b_1,\dots,b_n)$ or $\pm 1$\footnote{In all computations we have not seen the $-1$ show up.}. This structure is illustrated by $v_{1,0}(b_1,b_2)$ in~\eqref{v(1,2,0)}. To get every term in the volumes to have the structure~\eqref{totalstructure1}, the stuffle algebra of multi polylogs has to be used \cite{Waldschmidt2002}. This is because there will be products of multi polylogs that can be written in terms of linear combinations of multi polylogs. An example of this can be seen in the computation of~\eqref{v(1,2,0)} in Appendix \ref{details3}.

\section{Universal spectral form factor}
\label{section3}
The aim of this section is to compute the late time limit of the spectral form factor in unorientable JT gravity, and then to compute the spectral form factor from universal RMT with orthogonal symmetry. We find that the two expressions agree.
\subsection{Unorientable JT gravity}
\label{section3a}
In section \ref{background}, we introduced the $\tau$-scaled SFF
\begin{align}
  \kappa(\tau,\beta) \coloneqq \lim_{t \to \infty}\sum_{g=0,\frac{1}{2},1 \dots}^{\infty}\kappa_g(t,\beta)\tau^{2g+1}.
\end{align}
where
\begin{align}
  \kappa_g(t,\beta) \coloneqq \frac{Z^{(p)}_g(\beta+it,\beta-it)}{t^{2g+1}},
\end{align}
and $Z^{(p)}_g(\beta+it,\beta-it)$ can be computed from the double trumpet integral defined in~\eqref{p}. 
We will compute this expression up to $g=1$ with the volumes computed in section \ref{section2}.

The $g=0$ contribution is essentially trivial since $R_0(z_1,z_2)$ only depends on the support of the spectral curve and the result is the same as for the unorientable Airy model \cite{Weber2024}:
\begin{align}
\kappa_0(t,\beta) = \frac{1}{2\pi\beta} .
\end{align}
The $g=\frac{1}{2}$ contribution can be computed with the value of $V_\frac{1}{2}^{(p)}(b_1,b_2)$ given by~\eqref{1}
\begin{align}
   V^{(p)}_{\frac{1}{2}}(b_1,b_2) &= 4\log(p) + b_1+ 2\log(1+e^{\frac{b_2-b_1}{2}})+2\log(1+e^{-\frac{b_2+b_1}{2}}).
\end{align}
It is convenient for solving the double trumpet integral to write the volume in a manifestly symmetric form:
\begin{align}
   V^{(p)}_{\frac{1}{2}}(b_1,b_2)= V^{(p)}_{\frac{1}{2}}(b_1,b_2)\theta(b_1-b_2) +V^{(p)}_{\frac{1}{2}}(b_2,b_1)\theta(b_2-b_1),
\end{align}
which follows from the $b_1 \leftrightarrow b_2$ symmetry of the volumes. The $g=\frac{1}{2}$ coefficient of the $\tau$-scaled SFF is then
\begin{equation}
    \begin{aligned}
        \label{k1/2}
  \kappa_\frac{1}{2}(t,\beta) &=  \frac{1}{t^2}\int_0^\infty \int_0^\infty b_1\dd{b_1} b_2\dd{b_2}Z^t(b_1,\beta_1) Z^t(b_2,\beta_2)V^{(p)}_{\frac{1}{2}}(b_1,b_2)\theta(b_1-b_2)\\ 
  &+\left(\beta_1 \leftrightarrow \beta_2\right),
    \end{aligned}
\end{equation}
with $\beta_1 = \beta + it$, $\beta_2=\beta-it$, and $(\beta_1 \leftrightarrow \beta_2)$ implies copying the first term but switching  $\beta_1$ and $\beta_2$. The details of the integrals are relegated to Appendix \ref{g1/2}, but it is found that only the subset of terms,
\begin{align}
  v_{\frac 1 2,0}(b_1,b_2) \supset b_1+ 2\log(1+e^{\frac{b_2-b_1}{2}}),
\end{align}
contribute in the late time limit. We note that the divergent part of the volume $v_{\frac{1}{2},1}(b_1,b_2)$ gives contributions that are subleading in $t$ and does not contribute. Integrating this subset of terms in~\eqref{k1/2} results in:
\begin{equation}
    \begin{aligned}
        \label{k1/22}
  \kappa_{\frac{1}{2}}(t,\beta) =&  -\frac{1}{\sqrt{2\pi \beta}}-\frac{2}{\sqrt{2\pi \beta}} \sum_{k=1}^{\infty} (-1)^k \left(1-k\sqrt{\frac{\beta \pi}{2}} e^{\frac{\beta k^2}{2}} \text{erfc}\left(k\sqrt{\frac{\beta}{2}}\right)\right) \\ 
 &+ \order{t^{-\frac{1}{2}}}.
    \end{aligned}
\end{equation}
The first term is just the unorientable Airy term that was computed in \cite{Weber2024}. Considering the full theory of unorientable JT gravity leads to an additional infinity of terms, which is a reflection of the fact that in unorientable JT gravity one has to consider the infinity of zeros in the spectral curve. 

The contribution of $g=1$ is found in the same way by integrating $V^{(p)}_{1}(b_1,b_2)$ against the double trumpet. The only terms that can contribute in the $\tau-$scaled limit are terms at least quadratic in the lengths, e.g. $b_1^2, b_2^2$, $b_1b_2$, and terms such as $b_1e^{\frac{k}{2}(b_2-b_1)}$ where $k$ is a summation index. Therefore, all three volumes, $v_{1,2}(b_1,b_2), \\ v_{1,1}(b_1,b_2),$ and $v_{1,0}(b_1,b_2)$ can in principle contribute to the SFF. The relevant terms of $v_{1,2}(b_1,b_2)$, found from~\eqref{v(1,2,2)}, are,
\begin{equation}
    v_{1,2}(b_1,b_2) \supset 2b_1^2+2b_2^2
\end{equation}
and the relevant terms of $v_{1,2}(b_1,b_2)$, found from~\eqref{v(1,2,1)}, are
\begin{equation}
  v_{1,1}(b_1,b_2)  \supset \frac{1}{3} \left(2 b_1^3+3 b_2^2 b_1+b_2^3\right) +8\left(b_1-b_2\right) \text{Li}_2\left(e^{\frac{1}{2} \left(b_2-b_1\right)}\right).
\end{equation}
When integrated against the double trumpet, in the same way that was done in~\eqref{k1/2}, both these contributions individually give zero (see Appendix \ref{g1}), which means the SFF will have no logarithmic divergence through genus one.
The relevant terms of $v_{1,0}(b_1,b_2)$ found from~\eqref{v(1,2,0)} are,
\begin{equation}
    \begin{aligned}
        v_{1,0}(b_1,b_2) &\supset 
\frac{1}{12} b_1 \left(b_2^3+4 \pi ^2 b_2\right)+\frac{7 b_1^4}{96}+\left(\frac{7 b_2^2}{48}+\frac{5 \pi ^2}{6}\right) b_1^2+\frac{b_2^4}{32}+\frac{b_2^2\pi^2}{2} \\[8pt]
&+\left(2 b_1^2 -2b_2b_1\right) \text{Li}_2\left(e^{\frac{1}{2} \left(b_2-b_1\right)}\right)+8b_1 \text{Li}_3\left(e^{\frac{1}{2} \left(b_2-b_1\right)}\right)\\[8pt]
&+8(b_2-b_1) \text{Li}_{2,1}\left(e^{\frac{1}{2} \left(b_2-b_1\right)},1\right).
   \label{universal}
    \end{aligned}
\end{equation}
$\kappa_{1}(t,\beta)$ is then found by integrating~\eqref{universal} against the double trumpet. The details are given in Appendix \ref{g1} and the result is
\begin{align}
\label{k10}
    \kappa_1(t,\beta) &= \frac{1}{\pi}\left(\frac{-10}{3}+\log(\frac{2t}{\beta})\right)+\frac{4}{\pi}\sum_{k=1}^{\infty}\left(-1+\frac{1}{2}\left(1+\beta k^2\right)e^{\frac{\beta k^2}{2}}E_1\left(\frac{\beta k^2}{2}\right)\right) \\ \nonumber
    &+\order{t^{-\frac{1}{2}}},
\end{align}
where $E_1(x)$ is the exponential integral. The result is again in the form of the Airy result plus an additional infinity of terms. The $\tau$-scaled SFF up to $g=1$ computed from unorientable JT gravity is then
\begin{equation}
    \begin{aligned}
        \kappa(\tau,\beta) & = \kappa_{\text{Airy}}(\tau,\beta)-\frac{2\tau^2}{\sqrt{2\pi \beta}} \sum_{k=1}^{\infty} (-1)^k \left(1-k\sqrt{\frac{\beta \pi}{2}} e^{\frac{\beta k^2}{2}} \text{erfc}\left(k\sqrt{\frac{\beta}{2}}\right)\right) \\[8pt]
  &+\frac{4\tau^3}{\pi}\sum_{k=1}^{\infty}\left(-1+\frac{1}{2}\left(1+\beta k^2\right)e^{\frac{\beta k^2}{2}}E_1\left(\frac{\beta k^2}{2}\right)\right)+\order{\tau^4},
  \label{SFF-jt}
    \end{aligned}
\end{equation}
where the Airy contribution is
\begin{align}
  \kappa_{\text{Airy}}(\tau,\beta) = \frac{\tau}{2\pi \beta}-\frac{\tau^2}{\sqrt{2\pi \beta}} +\frac{\tau^3}{\pi}\left(\frac{-10}{3}+\log(\frac{2t}{\beta})\right) + \order{\tau^4}.
  \label{label}
\end{align}
The $t$-dependent terms stemming from the unorientable Airy contributions were discussed extensively in \cite{Weber2024}. We note that the final result is finite, i.e. independent of $\log(p)$, and therefore does not depend on the regularization used. We will see in the next section that this is necessary for the result to agree with universal RMT. However, this was not guaranteed to be the case since both $v_{1,2}(b_1,b_2)$ and $v_{1,1}(b_1,b_2)$ have terms that could, in principle, contribute in the $\tau$-scaled limit. The fact that all the contributions vanish can be seen as an example for constraints on the divergent parts of the volumes deriving from the SFF having to match to the result from universal RMT.
\subsection{Universal RMT}
\label{section3b}
The $\tau-$scaled SFF from universal RMT is given by Laplace transform of the \textit{microcanonical} SFF \cite{Weber2024}:
\begin{align}
\label{eq18}
 \kappa^{\text{GOE}}(\tau,\beta) = \int_{0}^{\infty}\dd E e^{-2\beta E} \kappa_E^{\text{GOE}}(\tau),
 \end{align}
 with 
 \begin{align}
 \kappa_E^{\text{GOE}}(\tau) = \rho_0(E)\left(1-b^{\text{GOE}}\left(\frac{\tau}{2\pi \rho_0(E)}\right)\right)
\end{align}
where $b^{\text{GOE}}(x)$ is the form factor and we have used notation consistent with \cite{Mehta2004}. We use the notion "universal" in the sense that the functional form of $b^{\text{GOE}}(x)$ only depends on the symmetry class, though the $\tau-$scaled SFF will be dependent on the leading order density of eigenvalues, $\rho_0(E)$. We take the symmetry class to be orthogonal, i.e. GOE. The superscript GOE also serves to distinguish the computation from the unorientable JT gravity computation. The form factor for the GOE is given in \cite{Mehta2004}:
 \begin{align}
  b^{\text{GOE}}\left(\frac{\tau}{2\pi \rho_0(E)}\right)=\left\{
	\begin{array}{ll}
		1-\frac{\tau}{\pi \rho_0(E)}+\frac{\tau}{2\pi \rho_0(E)}\log\qty(1+\frac{\tau}{\pi\rho_0(E)}) & \mbox{if } \frac{\tau}{2\pi} \leq \rho_0(E),  \\
		-1+\frac{\tau}{2\pi\rho_0(E)}\log\qty(\frac{\frac{\tau}{\pi}+\rho_0(E)}{\frac{\tau}{\pi}-\rho_0(E)}) & \mbox{if } \frac{\tau}{2\pi} \geq \rho_0(E).
	\end{array}
\right.
\label{eq:cf}
\end{align}
To compare the results of~\eqref{eq18} to the computation of unorientable JT gravity we need to input the leading order energy density of JT gravity:
\begin{align}
  \rho_0(E) = \frac{1}{4 \pi^2}\sinh(2\pi \sqrt{E}).
  \label{rho}
\end{align}
Following the steps outlined in Appendix \ref{appendixd}, the $\tau-$scaled SFF can be written as
\begin{align}
   \kappa^{\text{GOE}}(\tau,\beta) =2\kappa^{\text{GUE}}(\tau,\beta) +\chi(\tau,\beta),
   \label{generalform}
\end{align}
where:
\begin{align}
    \chi(\tau,\beta) =
\frac{-\tau}{4\pi\beta}\sum_{n=1}^{\infty}\frac{(-2\tau)^n}{n!}\int_{\tau}^{\infty}dx\frac{\frac{d^n}{dx^n}e^{-\frac{2\beta}{4\pi^2}\left(\text{arcsinh}(2\pi x)\right)^2}}{x},
\label{chi}
\end{align}
and $\kappa^{\text{GUE}}(\tau,\beta)$ was computed in \cite{Saad2022}:
\begin{equation}
  \begin{aligned}
  2\kappa^{\text{GUE}}(\tau,\beta) &= \frac{2e^{\frac{\pi^2}{2\beta}}}{16 \sqrt{2\pi}\beta^{3/2}}\left(\text{Erf}\left(\frac{\frac{\beta}{\pi}\text{arcsinh}(2\pi \tau)+\pi}{\sqrt{2\beta}}\right)+\text{Erf}\left(\frac{\frac{\beta}{\pi}\text{arcsinh}(2\pi \tau)-\pi}{\sqrt{2\beta}}\right)\right) \\
  &=\frac{\tau}{2\pi \beta}-\frac{\tau^3}{3\pi} +\order{\tau^4}.
\end{aligned}
\end{equation}
The integral in~\eqref{chi} can be computed perturbatively in $\tau$. For example, the $n=3$ term will only contribute to $\tau^4$ and higher, so we can compute the contribution up to $\tau^3$ by considering only the $n=1$ and $n=2$ terms. The details of the derivation are in Appendix \ref{appendixd} where it is shown that:
\begin{equation}
\begin{aligned}
&\kappa^{\text{GOE}}(\tau,\beta) = 2\kappa^{\text{GUE}}(\tau,\beta)+\chi(\tau,\beta) \\[8pt]
&=\frac{\tau}{2\pi \beta} -\frac{\tau^3}{3\pi}+ \frac{-\tau}{4\pi\beta}\sum_{n=1}^{2}\frac{(-2\tau)^n}{n!}\int_{\tau}^{\infty}dx\frac{\frac{d^n}{dx^n}e^{-\frac{2\beta}{4\pi^2}\left(\text{arcsinh}(2\pi x)\right)^2}}{x}+\order{\tau^4}\\[8pt] 
&= \kappa_{\text{Airy}}^{\text{GOE}}(\tau,\beta)-\frac{2\tau^2}{\sqrt{2\pi \beta}} \sum_{k=1}^{\infty} (-1)^k \left(1-k\sqrt{\frac{\beta \pi}{2}} e^{\frac{\beta k^2}{2}} \text{erfc}\left(k\sqrt{\frac{\beta}{2}}\right)\right) \\[8pt] 
& +\frac{4 \tau^3}{\pi}\sum_{k=1}^{\infty}
   \left(-1+\frac{1}{2}e^{\frac{\beta k^2}{2}}(1+\beta k^2)E_1\left(\frac{\beta k^2}{2}\right)\right)+\order{\tau^4},
   \label{GOE}
\end{aligned} 
\end{equation}
with the Airy contribution given by:
\begin{align}
\label{87}
  \kappa_{\text{Airy}}^{\text{GOE}}(\tau,\beta) = \frac{\tau}{2\pi \beta}-\frac{\tau^2}{\sqrt{2\pi \beta}}+\frac{-\tau^3}{\pi}\left(\gamma+\log(2\beta \tau^2) + \frac{1}{3}\right) + \order{\tau^4}.
\end{align}
By comparing~\eqref{GOE} and~\eqref{SFF-jt} it is seen all of the new contributions from unorientable JT gravity, i.e. not the Airy terms, agree with universal RMT. The Airy terms were first computed in \cite{Weber2024}, and there it was shown how the $\tau^3$ coefficient of $\kappa_{\text{Airy}}^{\text{GOE}}(\tau,\beta)$ and $\kappa_{\text{Airy}}(\tau,\beta)$  could be made to agree by using asymptotic expansions of generalized hypergeometric functions. However, this required computing higher order terms in the loop equations than one would expect, and we refer the reader to \cite{Weber2024} for the details.
\section{Discussion}
The fidelity, at least to order $\tau^3$, of the late time limit of the SFF in unorientable JT gravity, i.e. JT gravity with time reversal symmetry, to universal RMT with orthogonal symmetry demonstrates chaos in the sense of the BGS conjecture \cite{Bohigas1984}. Computing correlation functions in unorientable JT gravity requires computing the unorientable volumes. The computation of the unorientable volumes presents difficulty, not only because they require regularization but because it involves computing an infinity of residues in the loop equations. For comparison, the computation of the orientable volumes through topological recursion only requires computing residues at the end points of the spectral curve. We overcome this difficulty by introducing streamlined formulas to compute the volumes for one and two boundaries, which greatly reduces the amount of work required. 

By computing the volumes through genus one, we were able to conjecture the multiple polylogarithmic structure of the unorientable volumes. The algebraic properties of multiple polylogarithms, and their many reductions, allow the volumes to be cast in a much simpler form than what would be initially expected. It is interesting to note that there is a long history of multi zeta values occurring in the computation of moduli space volumes \cite{Zagier1994}. We also see zeta values at odd integers which is a direct consequence of the unorientability of the surfaces, something that was discussed in \cite{Witten1991a}. Furthermore, multiple polylogarithms occur frequently in the computation of Feynman integrals in quantum field theory \cite{Duhr2014,Duhr2019} and in string theory \cite{Schlotterer:2012, Mafra:2022}. It would be interesting to prove the conjectured structure of the unorientable volumes at the same level of rigor as Mirzakhani's theorem for the orientable volumes. This would further demonstrate the deep connections between multiple polylogarithms and algebraic geometry that has already been a rich subject of mathematics \cite{Goncharov1998,Goncharov2001, brown2012}. 

The agreement between the $\tau-$scaled SFF computed from the volumes and the result computed from universal RMT implies constraints on the coefficients of the volumes. For example, the $\tau-$scaled SFF from universal RMT does not have any divergence akin to the divergences encountered in the unorientable volumes. In contrast, due to precisely these divergencies, unorientable JT gravity is a divergent theory, and it is not obvious a priori that the $\tau-$scaled SFF computed within this theory would be finite. Consequently, the matching of the two results, being an effect of random matrix universality in chaotic systems, implies certain constraints on the divergent parts of the unorientable volumes. Moreover, the SFF from universal RMT can not break the $\tau-$scaling, in the sense of having remaining terms dependent on $t$, by construction. If we exclude the Airy terms, the contributions computed through genus one in unorientable JT gravity also do not break the $\tau-$scaling. Thus, also in this case the matching of the two computations of the canonical SFF suggests constraints on the volumes akin to what has been studied in the orientable case \cite{Blommaert2022, Weber2022}. The Airy terms are subtle and do break the $\tau-$scaling, which leads to having $t-$dependent terms. In \cite{Weber2024} it was shown this issue could be resolved by utilizing a pseudo-renormalization scheme, involving asymptotic expansions of generalized hypergeometric functions.

A possible direction of future work would be to solve the loop equations using the super JT gravity spectral curve: $ y^{\text{SJT}}(z) \propto z^{-1}\cos(z) $. For super JT gravity the proper RMT ensemble would be generically one of the Altland-Zirnbauer symmetry classes \cite{Stanford2019, Altland1997}. The microcanonical form factor, necessary to compute the SFF from universal RMT for the non Wigner-Dyson symmetry classes is given e.g. in \cite{Gnutzmann2004}. Work in the sense of this program was already presented in \cite{Griguolo2024} for the variation of super JT gravity dual to a chiral GUE ensemble. There, methods of resurgent analysis were used on the gravity side that would also be expected to be necessary for variations of super JT gravity involving time-reversal invariance. Supersymmetric JT gravity and matrix models has also been explored in \cite{Turiaci2023}.

A rather direct use of the results found in this work would be to consider the $(2,2p+1)$ minimal string model, used here in the limit of large $p$ corresponding to JT gravity, for finite $p$. The minimal string models display a large amount of interesting mathematical properties in the orientable setting (cf. \cite{Eynard2016} for an overview). Computing the correlation functions for minimal string models with time-reversal invariance to our knowledge has not yet been performed. This would enable studying the theories in the unorientable setting and might lead to novel insights into their mathematical properties.

\section*{Acknowledgments}
We would like to thank F. Haneder, M. Lents, and S. Tomsovic for valuable discussions. We acknowledge financial support from the Deutsche Forschungsgemeinschaft (German Research Foundation) through Ri681/15-1 (project number 456449460) within the Reinhart-Koselleck Programme.

 \appendix
 \section{Details of volume computations}
 \label{details}
 \subsection{$v_{1,k}(b)$}
 \label{details1}
 The $(g,n)=(1,1)$ unorientable volume is computed by using~\eqref{eq:F11} in~\eqref{eq:v_11_0}:
 \begin{equation}
     \begin{aligned}
         V_1^{(p)}(b) 
   &=\left(\underset{z=0}{\Res}+\sum_{k=1}^{\infty}\underset{z=-\frac{k}{2}}{\Res}\right)\left(\frac{ ze^{bz}}{by(z)}\right)
  \left(\frac{1}{2z}\partial_{z}R^{(p)}_{\frac{1}{2}}(z) + R_{0}(z,z)+R^{(p)}_{\frac{1}{2}}(z)^2\right) \\ 
  &+ \order{b^{-1}}.
     \end{aligned}
 \end{equation}
 We will compute the volume in the notation of the expansion~\eqref{eq:regvol} and~\eqref{regres}. The coefficient of $\log(p)^2$, only comes from the term $R^{(p)}_{\frac{1}{2}}(z)^2$, i.e. $r^2_{\frac{1}{2},1}(z)$. The value of $r_{\frac{1}{2},1}(z)$ can be read off from~\eqref{eq:crosscapr}: $r_{\frac{1}{2},1}(z)= \frac{-1}{z}$. This results in the following:

\begin{equation}
    \begin{aligned}
         v_{1,2}(b)&= \left(\underset{z=0}{\Res}+\sum_{k=1}^{\infty}\underset{z=-\frac{k}{2}}{\Res}\right)\left(\frac{ ze^{bz}}{by(z)}\right) r^2_{1,1}(z) +\order{b^{-1}} \\
 &= \underset{z=0}{\Res}\left(\frac{2e^{bz}}{bz^2}\right) + \order{b^{-1}}\\
 &=2.
    \end{aligned}
\end{equation}
To get to the second line we used the expansion of the inverse spectral curve~\eqref{eq:expansion2} and in the third line we dropped the $\order{b^{-1}}$ because there cannot be any terms that go as $\order{b^{-1}}$ in the final result, and if we had kept every term of order $\order{b^{-1}}$ they would have all canceled. The computation of $v_{1,1}(b)$ is as follows:
\begin{align}
   v_{1,1}(b)&= \left(\underset{z=0}{\Res}+\sum_{k=1}^{\infty}\underset{z=-\frac{k}{2}}{\Res}\right)\left(\frac{ ze^{bz}}{by(z)}\right) \left(\frac{1}{2z}\partial_z r_{\frac{1}{2},1}(z) +2r_{\frac{1}{2},1}(z)r_{\frac{1}{2},0}(z)\right) +\order{b^{-1}}.
\end{align}
From~\eqref{eq:crosscapr} it is seen that:
\begin{align}
 r_{\frac{1}{2},0}(z) &= \frac{-1}{4z^2} + \frac{H_{2z}}{z} ,
\end{align}
where: 
\begin{align}
  H_{2z}=\sum^{\infty}_{k=1}\left(\frac{1}{k}-\frac{1}{1+2z}\right).
  \label{expansion3}
\end{align}
The residue at $z=0$ is:

\begin{equation}
    \begin{aligned}
        v_{1,1}(b) &\supset \underset{z=0}{\Res}\left(\frac{ 2e^{bz}}{b}\right) \left(\frac{1}{2z^3} +\frac{1}{2z^3}-\frac{2H_{2z}}{z^2}\right) +\order{b^{-1}} \\[10pt]
  &=b +\order{b^{-1}}.
  \label{term1}
    \end{aligned}
\end{equation}
The residue at the negative half integers can only contribute from the series in the expansion of $y(z)^{-1}$,~\eqref{eq:expansion2}, and $H_{2z}$,~\eqref{expansion3}, when the indices of the two sums are equal:
\begin{equation}
    \begin{aligned}
         v_{1,1}(b) &\supset \sum_{k_1=1}^{\infty}\underset{z=-\frac{k_1}{2}}{\Res}\sum^{\infty}_{k_2=1}\sum^{\infty}_{k_3=1}\frac{-8e^{bz}(-1)^{k_2}}{b(z-\frac{k_2}{2})(z+\frac{k_2}{2})}\left(\frac{1}{k_3}-\frac{1}{k_3+2z}\right)+\order{b^{-1}}\\[10pt]
  &= \sum_{k=1}^{\infty}\frac{-4(-1)^k e^{-\frac{bk}{2}}}{k} +\order{b^{-1}} \\[10pt]
  &=4\log(1+e^{-\frac{b}{2}})+ \order{b^{-1}}.
  \label{term2}
    \end{aligned}
\end{equation}
In the second line we took the residue when $k_2=k_3$. Combining~\eqref{term1} and~\eqref{term2} then gives:
\begin{equation}
    \begin{aligned}
        v_{1,1}(b) &= b+4\log(1+e^{-\frac{b}{2}}) \\
  &=4\log(2\cosh(\frac{b}{4})),
    \end{aligned}
\end{equation}
where we have again dropped the $\order{b^{-1}}$ since it cannot be in the final result. Moving on to $v_{1,0}(b)$, and plugging in the value of $R_0(z,z)$ given by~\eqref{special}:
\begin{equation}
    \begin{aligned}
        v_{1,0}(b) &= \\
   &\left(\underset{z=0}{\Res}+\sum_{k=1}^{\infty}\underset{z=-\frac{k}{2}}{\Res}\right)\left(\frac{ ze^{bz}}{by(z)}\right)
  \left(\frac{1}{2z}\partial_{z}r_{\frac{1}{2},0}(z) + \frac{1}{8z^4}+r_{\frac{1}{2},0}(z)^2\right) \\
  &+ \order{b^{-1}} .
  \label{eq:v11_2a}
    \end{aligned}
\end{equation}
The residue at $z=0$ is straightforward, using the expansion~\eqref{eq:expansion2}:
\begin{equation}
    \begin{aligned}
        v_{1,0}(b) &\supset \underset{z=0}{\Res}\left(\frac{ ze^{bz}}{by(z)}\right)
  \left(\frac{1}{2z}\partial_{z}r_{\frac{1}{2},0}(z) + \frac{1}{8z^4}+r_{\frac{1}{2},0}(z)^2\right)+ \order{b^{-1}} \\[10pt] 
  & = \frac{7b^2}{48} + \frac{\pi^2}{4}+\order{b^{-1}}.
  \label{term3}
    \end{aligned}
\end{equation}
The residues at $z=-\frac{k}{2}$ are slightly more involved:
\begin{equation}
    \begin{aligned}
        v_{1,0}(b) &\supset \sum_{k=1}^{\infty}\underset{z=-\frac{k}{2}}{\Res}\left(\frac{ ze^{bz}}{by(z)}\right)
  \left(\frac{1}{2z}\partial_{z}r_{\frac{1}{2},0}(z)+r_{\frac{1}{2},0}(z)^2\right) + \order{b^{-1}} \\[10pt]
  &=\sum_{k=1}^{\infty}\underset{z=-\frac{k}{2}}{\Res}\left(\frac{ e^{bz}}{b}\right)
  \left(-\frac{1}{2}\left(\frac{b}{y(z)}-\frac{\partial_{z}y(z)}{y(z)^2}\right)r_{\frac{1}{2},0}(z) +z\frac{r_{\frac{1}{2},0}(z)^2}{y(z)}\right) + \order{b^{-1}}.
  \label{v10}
    \end{aligned}
\end{equation}
In the first line we dropped the contribution from $R_0(z,z)$ and in the second line an integration by parts was used. The first term can be computed by noting that:
\begin{align}
  \frac{\partial_{z}y(z)}{y(z)^2} = \frac{8\pi^2\cos(2\pi z)}{\sin(2\pi z)^2}=2\cos(2\pi z)\left(\frac{1}{z^2}+\sum_{k=1}^{\infty}\left(\frac{1}{(z+\frac{k}{2})^2}+\frac{1}{(z-\frac{k}{2})^2}\right)\right),
\end{align}
and only the term with a double pole at $z=\frac{-k}{2}$ will be relevant. Using this expression to evaluate the first term in~\eqref{v10} we find:
\begin{align}
   v_{1,0}(b) &\supset
  \sum_{k=1}^{\infty}\underset{z=-\frac{k}{2}}{\Res}\left(\frac{ e^{bz}}{b}\right)
  \left(-\frac{1}{2}\left(\frac{b}{y(z)}-\sum_{k_1=1}^{\infty}\frac{2\cos(2\pi z)}{(z+\frac{k_1}{2})^2}\right)\left(\frac{-1}{4z^2}+\frac{1}{z}H_{2z}\right)\right) + \order{b^{-1}}.
\end{align}
 The residues again can be computed using~\eqref{eq:expansion2} and~\eqref{expansion3} while taking care to separate out terms when the summation indices are equal, i.e. $k_1=k_2$. The result is: 
\begin{equation}
    \begin{aligned}
        \label{v102}
   v_{1,0}(b) &\supset \sum_{k=1}^{\infty}\underset{k_1 \neq k}{\sum_{k_1=1}^{\infty}}\frac{-4 (-1)^k e^{-\frac{1}{2} \left(b k_1\right)}}{k^2-k_1^2}+ 2\text{Li}_2\left(e^{-\frac{b}{2}}\right)+\text{Li}_2\left(-e^{-\frac{b}{2}}\right)+\frac{b}{2} \log \left(e^{-\frac{b}{2}}+1\right) \\ 
   &+ \order{b^{-1}}.
    \end{aligned}
\end{equation}
 The second term in~\eqref{v10} is:
 \begin{equation}
    \begin{aligned}
   v_{1,0}(b) &\supset
 \sum_{k=1}^{\infty}\underset{z=-\frac{k}{2}}{\Res}\left(\frac{ e^{bz}}{b}\right)
  \frac{z}{y(z)}\left(\frac{-1}{4z^2}+\frac{1}{z}H_{2z}\right)^2 + \order{b^{-1}} \\
  &=\sum_{k=1}^{\infty}\underset{z=-\frac{k}{2}}{\Res}\left(\frac{ e^{bz}}{b}\right)
  \frac{z}{y(z)}\left(-\frac{1}{2z^3}H_{2z}+\frac{1}{z^2}H_{2z}^2\right) + \order{b^{-1}} \\[8pt] 
  \label{v103}
  &= 2\text{Li}_2\left(e^{-\frac{b}{2}}\right)+\text{Li}_2\left(-e^{-\frac{b}{2}}\right)+\frac{b}{2} \log \left(e^{-\frac{b}{2}}+1\right)+ 2\log(e^{-\frac b 2}+1)^2 \\ 
  & + \sum_{k=1}^{\infty}\underset{k_1 \neq k}{\sum_{k_1=1}^{\infty}}\frac{-4 (-1)^k e^{-\frac{1}{2} \left(b k_1\right)}}{k^2-k_1^2}+\order{b^{-1}}.
    \end{aligned}
\end{equation}
 Note we used polylogarithm identies to get this result \cite{maximon2003}. The sum can be written in terms of multi-polylogarithms:
 \begin{align}
 \label{a16}
  & \sum_{k=1}^{\infty}\underset{k_1 \neq k}{\sum_{k_1=1}^{\infty}}\frac{- (-1)^k e^{-\frac{1}{2} \left(b k_1\right)}}{k^2-k_1^2}  = \\ \nonumber
  &\frac{1}{2}\text{Li}_{1,1}\left(-1,-e^{-\frac{b}{2}}\right)+\frac{1}{2}\text{Li}_{1,1}\left(-e^{-\frac{b}{2}},-1\right)-\frac{1}{4}\text{Li}_2\left(-e^{-\frac{b}{2}}\right)-\frac{1}{2}\text{Li}_1\left(-e^{-\frac{b}{2}}\right)\text{Li}_{1}\left(-1\right) \\[8pt] \nonumber
  &=-\frac{1}{2}\text{Li}_2\left(e^{-\frac{b}{2}}\right)-\frac{1}{4}\text{Li}_2\left(-e^{-\frac{b}{2}}\right)
 \end{align}
 To get the last line we used a stuffle identity \cite{Waldschmidt2002}.
 The final result is found by combining~\eqref{term3},~\eqref{v102},~\eqref{v103}, and using~\eqref{a16}:
 \begin{equation}
 \begin{aligned}
 \label{eq:v10}
  v_{1,0}(b) =\frac{7b^2}{48} +\frac{\pi^2}{4}+b\log\left(1+e^{-\frac{b}{2}}\right)
  + 2\log\left(1+e^{-\frac{b}{2}}\right)^2.
 \end{aligned}
 \end{equation}
 \subsection{$v_{\frac{1}{2},k}(b_1,b_2)$}
 \label{details2}
 The computation of the $(g,n)=(\frac{1}{2},2)$ unorientable volume follows by plugging~\eqref{F1/2} into~\eqref{v1/2}:
 \begin{equation}
     \begin{aligned}
         \label{v1/2_1}
  V^{(p)}_{\frac{1}{2}}(b_1,b_2) &=\left(\underset{z_2=0}{\Res}\hspace{.1cm}\underset{z_1=0}{\Res}+\sum_{k=1}^{\infty}\underset{z_2=\pm \frac{k}{2}}{\Res}\underset{z_1=-\frac{k}{2}}{\Res}\right)\frac{-4z_1z_2 }{2y(z_1)b_1 b_2}e^{b_1z_1}e^{b_2 z_2}  \\
  &\times\left(\frac{1}{2z_1}\partial_{z_1} R_{0}(z_1,z_2) +2R^{(p)}_{\frac{1}{2}}(z_1)\left(R_{0}(z_1,z_2)+\frac{1}{(z_1^2-z_2^2)^2}\right)\right)\\
  &+\order{b_1^{-1}}+\order{b_2^{-1}}.
     \end{aligned}
 \end{equation}
The only contribution to $v_{\frac{1}{2},1}(b_1,b_2)$ comes from $r_{\frac{1}{2},1}(z_1)$ = $\frac{-1}{z_1}$:
\begin{equation}
    \begin{aligned}
        v_{\frac{1}{2},1}(b_1,b_2)&=\underset{z_2=0}{\Res}\hspace{.1cm}\underset{z_1=0}{\Res}\frac{4z_2 }{y(z_1)b_1 b_2}e^{b_1z_1}e^{b_2 z_2}\left(R_{0}(z_1,z_2)+\frac{1}{(z_1^2-z_2^2)^2}\right) \\
  &+\order{b_1^{-1}}+\order{b_2^{-1}} \\
  &=4
    \end{aligned}
\end{equation}
The computation of $v_{\frac 1 2, 0}(b_1,b_2)$ proceeds as follows:
\begin{equation}
    \begin{aligned}
        v_{\frac{1}{2},0}(b_1,b_2) &=\left(\underset{z_2=0}{\Res}\hspace{.1cm}\underset{z_1=0}{\Res}+\sum_{k=1}^{\infty}\underset{z_2=\pm \frac{k}{2}}{\Res}\underset{z_1=-\frac{k}{2}}{\Res}\right)\frac{-4z_1z_2 }{2y(z_1)b_1 b_2}e^{b_1z_1}e^{b_2 z_2}  \\
  &\times\left(\frac{1}{2z_1}\partial_{z_1} R_{0}(z_1,z_2) +2\left(-\frac{1}{4z_1^2}+\frac{1}{z_1}H_{2z_1}\right)\left(R_{0}(z_1,z_2)+\frac{1}{(z_1^2-z_2^2)^2}\right)\right)\\
  &+\order{b_1^{-1}}+\order{b_2^{-1}}.
    \end{aligned}
\end{equation}
The residues at zero can be computed by replacing $y(z_1)^{-1} \to \frac{2}{z_1}$ and by dropping $H_{2z_1}$, since the series in $y(z_1)^{-1}$ will only produce terms as $\order{b_1^{-1}}$ and $H_{0}=0$:
\begin{equation}
    \begin{aligned}
        v_{\frac{1}{2},0}(b_1,b_2) &\supset \underset{z_2=0}{\Res}\hspace{.1cm}\underset{z_1=0}{\Res} \frac{-4z_2 }{b_1 b_2}e^{b_1z_1}e^{b_2 z_2}  \\
  &\times\left(\frac{1}{2z_1}\partial_{z_1} R_{0}(z_1,z_2) +2\left(\frac{-1}{4z_1^2}\right)\left(R_{0}(z_1,z_2)+\frac{1}{(z_1^2-z_2^2)^2}\right)\right)\\
  &+\order{b_1^{-1}}+\order{b_2^{-1}} \\[8pt]
  & =b_1.
    \end{aligned}
\end{equation}
The only term that can contribute to the residues at $z_1=-\frac{k}{2}$ are the sums in the $H_{2z_1}$ and $y(z_1)^{-1}$ when the indices are equal:
\begin{equation}
    \begin{aligned}
        v_{\frac{1}{2},0}(b_1,b_2) &\supset \sum_{k=1}^{\infty}\underset{z_2=\pm \frac{k}{2}}{\Res}\underset{z_1=-\frac{k}{2}}{\Res}\frac{-4z_2 }{b_1 b_2}e^{b_1z_1}e^{b_2 z_2}  \\
  &\times\left(\sum_{k_1=1}^{\infty}\frac{4(-1)^kz_1}{(z_1-\frac{k_1}{2})(z_1+\frac{k_1}{2})}\left(\frac{1}{k_1}-\frac{1}{k_1+2z_1}\right)\right)\left(R_{0}(z_1,z_2)+\frac{1}{(z_1^2-z_2^2)^2}\right)\\
  &+\order{b_1^{-1}}+\order{b_2^{-1}}.  \\[8pt]
  & = 2\log(1+e^{\frac{b_2-b_1}{2}})+2\log(1+e^{-\frac{b_2-b_1}{2}}).
    \end{aligned}
\end{equation}
To get the last line we summed over the residues at $z_2=+\frac{k}{2}$ and $z_2 = -\frac{k}{2}$.
\subsection{$v_{1,k}(b_1,b_2)$}
\label{details3}
We can compute the $(g,n,k)=(1,2,k)$ unorientable volume using,
\begin{equation}
    \begin{aligned}
  \label{v_{1,0}1}
   &v_{1,k}(b_1,b_2) = \\[8pt]&\left(\sum_{k_1=1}^{\infty}\underset{z_2=\frac{k_1}{2}}{\Res}\hspace{.1cm}\underset{z_1=-\frac{k_1}{2}}{\Res}+\sum_{k_2=0}^{\infty}\sum_{k_1=0}^{\infty}\underset{z_2=-\frac{k_2}{2}}{\Res}\underset{z_1=-\frac{k_1}{2}}{\Res}\right) -\frac{4z_1z_2f_{1,k}(z_1,z_2)e^{b_1 z_2}e^{b_2 z_2}}{2b_1 b_2y(z_1)} \\[8pt]
   &+\order{b_1^{-1}}+\order{b_2^{-1}},
\end{aligned}
\end{equation}
The $f_{1,k}(z_1,z_2)$ can be found from the following formulas:
\begin{equation}
    \begin{aligned}
     &   F^{(p)}_1(z_1,z_2)=\frac{1}{2z_1}\partial_{z_1}R^{(p)}_{\frac{1}{2}}(z_1,z_2) + R_{0}(z_1,z_1,z_2) +\\[8pt]
  &2R^{(p)}_1(z_1)\left(\frac{1}{2}\frac{1}{z_1 z_2 \qty(z_1+z_2)^2}+\frac{1}{(z_1^2-z_2^2)^2}\right) +2R^{(p)}_{\frac{1}{2}}(z_1)R^{(p)}_{\frac{1}{2}}(z_1,z_2),
\\[8pt]
   &R^{(p)}_{\frac{1}{2}}(z_1,z_2) = \frac{-F^{(p)}_{\frac{1}{2}}(z_1,z_2)}{2y(z_1)} +\sum_{k=1}^{\infty}\frac{2(-1)^k\left(\frac{k}{2}\right)^2F^{(p)}_{\frac{1}{2}}(\frac{k}{2},z_2)}{z_1(z_1-\frac{k}{2})(z_1+\frac{k}{2})}+ \\
  \nonumber &\frac{1}{y(z_2)}R^{(p)}_{\frac{1}{2}}(z_2)\left(R_{0}(z_1,z_2)+\frac{1}{(z_1^2-z_2^2)^2}\right)-\frac{1}{2z_1(z_2^2-z_1^2)}\left(\partial_{z_2}\frac{R^{(p)}_{\frac{1}{2}}(z_2)}{y(z_2)}\right),
\\[8pt]
&        R^{(p)}_1(z)
= -\frac{F_1^{(p)}(z)}{2y(z)}+\sum_{k=1}^{\infty}\frac{2(-1)^k\left(\frac{k}{2}\right)^2 F_1^{(p)}(\frac{k}{2})}{z(z-\frac{k}{2})(z+\frac{k}{2})}.
\\[8pt]
  &F^{(p)}_{\frac{1}{2}}(z_1,z_2) =\frac{1}{2z_1}\partial_{z_1} R_{0}(z_1,z_2) +2R^{(p)}_{\frac{1}{2}}(z_1)\left(R_{0}(z_1,z_2)+\frac{1}{(z_1^2-z_2^2)^2}\right),
\\[8pt]
 &F_1^{(p)}(z) = \frac{1}{2z}\partial_{z}R^{(p)}_{\frac{1}{2}}(z) + R_{0}(z,z)+R^{(p)}_{\frac{1}{2}}(z)^2 \\[8pt]
 & R^{(p)}_{\frac{1}{2}}(z) = \frac{-1}{4z^2}+\frac{1}{z}H_{2z}-\frac{1}{z}\log(p)+\order{p^{-1}},
    \end{aligned}
\end{equation}
with
\begin{align}
&F^{(p)}_g(z_1,\dots,z_n) = \sum_{k=0}^{2g}f_{g,k}(z_1,\dots,z_n)\log(p)^k\\[8pt]
    &R^{(p)}_g(z_1,\dots,z_n) = \sum_{k=0}^{2g}r_{g,k}(z_1,\dots,z_n)\log(p)^k.
\end{align}
We note that some things simplify greatly, such as,
\begin{align}
    r_{1,2}(z_1) = -\frac{1}{z_1^3}, \hspace{.2cm} r_{\frac{1}{2},1}(z_1,z_2)=\frac{1}{z_1^3z_2^3},\hspace{.2cm}r_{\frac{1}{2},0}(z_1)=-\frac{-1}{4z_1^2}.
\end{align}
\paragraph{k=2} The easiest volume to compute is when $k=2$. The necessary formula is,
\begin{equation}
    \begin{aligned}
    f_{1,2}(z_1,z_2)=
  &2r_{1,2}(z_1)\left(\frac{1}{2}\frac{1}{z_1 z_2 \qty(z_1+z_2)^2}+\frac{1}{(z_1^2-z_2^2)^2}\right) +2r_{\frac{1}{2},1}(z_1)r_{\frac{1}{2},1}(z_1,z_2).
    \end{aligned}
\end{equation}
We find that only the residues at zero will contribute so that:
\begin{align}
    v_{1,2}(b_1,b_2) &= \underset{z_2=0}{\Res}\hspace{.1cm}\underset{z_1=0}{\Res}  -\frac{4z_1z_2f_{1,2}(z_1,z_2)e^{b_1 z_2}e^{b_2 z_2}}{2b_1 b_2y(z_1)} + \order{b_1^{-1}}+\order{b_2^{-1}}\\&=2b_1^2+2b_2^2+8\pi^2
\end{align}
\paragraph{k=1} Moving on to $v_{1,1}(b_1,b_2)$:
\begin{align}
 v_{1,1}(b_1,b_2) & \supset \underset{z_2=0}{\Res}\hspace{.1cm}\underset{z_1=0}{\Res} -\frac{4z_1z_2f_{1,1}(z_1,z_2)e^{b_1 z_2}e^{b_2 z_2}}{2b_1 b_2y(z_1)} + \order{b_1^{-1}}+\order{b_2^{-1}}\\
 & =\frac{1}{3} \left(2 b_1^3+3 b_2^2 b_1+b_2^3+4 \pi ^2 \left(3 b_1+b_2\right)+168 \zeta (3)\right).
\end{align}
The next term is:
\begin{align}
     v_{1,1}(b_1,b_2) & \supset \sum_{k=1}^{\infty}\underset{z_2=-k/2}{\Res}\hspace{.1cm}\underset{z_1=0}{\Res}-\frac{4z_1z_2f_{1,1}(z_1,z_2)e^{b_1 z_2}e^{b_2 z_2}}{2b_1 b_2y(z_1)}+ \order{b_1^{-1}}+\order{b_2^{-1}} \\
     &=8\text{Li}_3\left(e^{-b_2}\right)+64\sum_{k_2=1}^{\infty}\underset{k_1 \neq k_2}{\sum_{k_1=1}^{\infty}}\frac{\left((-1)^{k_1}+(-1)^{k_2}\right) e^{-\frac{1}{2} b_2 k_1}}{k_1\left(k_1^2- k_2^2\right)}.
\end{align}
We next consider:
\begin{align}
      v_{1,1}(b_1,b_2) & \supset \underset{z_2=0}{\Res}\hspace{.1cm}\sum_{k=1}^{\infty}\underset{z_1=-\frac{k}{2}}{\Res}-\frac{4z_1z_2f_{1,1}(z_1,z_2)e^{b_1 z_2}e^{b_2 z_2}}{2b_1 b_2y(z_1)}+ \order{b_1^{-1}}+\order{b_2^{-1}}\\
      & = \underset{z_2=0}{\Res}\hspace{.1cm}\sum_{k=1}^{\infty}\underset{z_1=-\frac{k}{2}}{\Res}-\frac{8z_1z_2e^{b_1 z_2}e^{b_2 z_2}}{2b_1 b_2y(z_1)}\left(r_{\frac{1}{2},0}(z_1)r_{\frac{1}{2},1}(z_1,z_2) +r_{\frac{1}{2},1}(z_1)r_{\frac{1}{2},0}(z_1,z_2) \right) \\
      &= -32\text{Li}_3\left(-e^{-\frac{b_1}{2}}\right).
\end{align}
The final term is:
\begin{align}
      v_{1,1}(b_1,b_2) & \supset \sum_{k=1}^{\infty}\left(\underset{z_2=\frac{k}{2}}{\Res}+\underset{z_2=-\frac{k}{2}}{\Res}\right)\hspace{.1cm}\underset{z_1=-\frac{k}{2}}{\Res}-\frac{4z_1z_2f_{1,1}(z_1,z_2)e^{b_1 z_2}e^{b_2 z_2}}{2b_1 b_2y(z_1)}\\[8pt] \nonumber
      & + \order{b_1^{-1}}+\order{b_2^{-1}} \\[8pt]
      &=\sum_{k\neq k_1}-\frac{64 \left(e^{-\frac{1}{2} \left(b_1+b_2\right) k} +e^{-\frac{1}{2} \left(b_1-b_2\right) k}\right)}{k \left(k-k_1\right) \left(k+k_1\right)}-16 \left(\text{Li}_3\left(e^{-\frac{b_1}{2}-\frac{b_2}{2}}\right)+\text{Li}_3\left(e^{\frac{1}{2} \left(b_2-b_1\right)}\right)\right)\\[8pt] \nonumber
      &+8\left(b_1-b_2\right) \text{Li}_2\left(e^{\frac{1}{2} \left(b_2-b_1\right)}\right)+8\left(b_1+b_2\right) \text{Li}_2\left(e^{-\frac{b_1}{2}-\frac{b_2}{2}}\right).
\end{align}
Adding everything together gives:
\begin{align}
   & v_{1,1}(b_1,b_2)=\frac{1}{3} \left(2 b_1^3+3 b_2^2 b_1+b_2^3+4 \pi ^2 \left(3 b_1+b_2\right)+168 \zeta (3)\right) \\[8pt] \nonumber
    &+32\text{Li}_3\left(e^{-\frac{b_2}{2}}\right)+32\text{Li}_3\left(-e^{-\frac{b_2}{2}}\right)-32\text{Li}_3\left(-e^{-\frac{b_1}{2}}\right)-16 \left(\text{Li}_3\left(e^{-\frac{b_1}{2}-\frac{b_2}{2}}\right)+\text{Li}_3\left(e^{\frac{1}{2} \left(b_2-b_1\right)}\right)\right)\\[8pt] \nonumber
    &+8\left(b_1-b_2\right) \text{Li}_2\left(e^{\frac{1}{2} \left(b_2-b_1\right)}\right)+8\left(b_1+b_2\right) \text{Li}_2\left(e^{-\frac{b_1}{2}-\frac{b_2}{2}}\right)\\[8pt] \nonumber &+64\sum_{k_1=1}^{\infty}\underset{k \neq k_1}{\sum_{k=1}^{\infty}}\frac{\left((-1)^{k}+(-1)^{k_1}\right) e^{-\frac{1}{2} b_2 k}}{k\left(k^2- k_1^2\right)}-\frac{\left(e^{-\frac{1}{2} \left(b_1+b_2\right) k} +e^{-\frac{1}{2} \left(b_1-b_2\right) k}\right)}{k \left(k^2-k_1^2\right) }.
\end{align}
The sums can be evaluated:
\begin{equation}
\begin{aligned}
   & \sum_{k_1=1}^{\infty}\underset{k \neq k_1}{\sum_{k=1}^{\infty}}\frac{\left((-1)^{k_1}\right) e^{-\frac{1}{2} b_2 k}}{k\left(k^2- k_1^2\right)}=\\
   &\frac{1}{2}\left(\text{Li}_{1,2}\left(-1,-e^{-\frac{1}{2}b_2}\right)+\text{Li}_{2,1}\left(-e^{-\frac{1}{2}b_2},-1\right)-\frac{1}{2}\text{Li}_3\left(-e^{-\frac{1}{2}b_2}\right) - \text{Li}_2\left(-e^{-\frac{1}{2}b_2}\right)\text{Li}_1\left(-1\right)\right) \\
   &=-\frac{3}{4}\text{Li}_3\left(-e^{-\frac{1}{2}b_2}\right),
\end{aligned}
\end{equation}
where we used the following identity:
\begin{equation}
\begin{aligned}
  \text{Li}_{1,2}\left(-1,-e^{-\frac{1}{2}b_2}\right)+\text{Li}_{2,1}\left(-e^{-\frac{1}{2}b_2},-1\right) - \text{Li}_2\left(-e^{-\frac{1}{2}b_2}\right)\text{Li}_1\left(-1\right)=-\text{Li}_3\left(-e^{-\frac{1}{2}b_2}\right).
\end{aligned}
\end{equation}
The next sum is
\begin{equation}
\begin{aligned}
 &   \sum_{k_1=1}^{\infty}\underset{k \neq k_1}{\sum_{k=1}^{\infty}}\frac{\left((-e^{-\frac{b_2}{2}})^k-e^{-\frac{1}{2} \left(b_1+b_2\right) k} -e^{-\frac{1}{2} \left(b_1-b_2\right) k}\right)}{k \left(k^2-k_1^2\right) }\\
 &-\frac{3}{4}\left(\text{Li}_3\left(-e^{-\frac{1}{2}b_2}\right)-\text{Li}_3\left(e^{-\frac{1}{2}\left(b_1+b_2\right)}\right)-\text{Li}_3\left(e^{-\frac{1}{2}\left(b_1-b_2\right)}\right)\right).
\end{aligned}
\end{equation}
The final result is then:
\begin{align}
   & v_{1,1}(b_1,b_2)=\frac{1}{3} \left(2 b_1^3+3 b_2^2 b_1+b_2^3+4 \pi ^2 \left(3 b_1+b_2\right)+168 \zeta (3)\right) \\[8pt] \nonumber
    &-32\text{Li}_3\left(-e^{-\frac{b_2}{2}}\right)-32\text{Li}_3\left(-e^{-\frac{b_1}{2}}\right)+32\text{Li}_3\left(e^{-\frac{b_1}{2}-\frac{b_2}{2}}\right)+32\text{Li}_3\left(e^{\frac{1}{2} \left(b_2-b_1\right)}\right)\\[8pt] \nonumber
    &+8\left(b_1-b_2\right) \text{Li}_2\left(e^{\frac{1}{2} \left(b_2-b_1\right)}\right)+8\left(b_1+b_2\right) \text{Li}_2\left(e^{-\frac{b_1}{2}-\frac{b_2}{2}}\right).
\end{align}
We have numerically verified the symmetry of this result with respect to $(b_1 \leftrightarrow b_2)$. 
\paragraph{k=0} To compute $v_{1,0}(b_1,b_2)$ we need,
\begin{align}
\nonumber
   f_{1,0}(z_1,z_2)=& \frac{1}{2z_1}\partial_{z_1}r_{\frac{1}{2},0}(z_1,z_2) + R_{0,3}(z_1,z_1,z_2) +\\
  &2r_{1,0}(z_1)\left(\frac{1}{2}\frac{1}{z_1 z_2 \qty(z_1+z_2)^2}+\frac{1}{(z_1^2-z_2^2)^2}\right) +2r_{\frac{1}{2},0}(z_1)r_{\frac{1}{2},0}(z_1,z_2).
  \label{f101}
\end{align}
The resolvents in the expression~\eqref{f101} can be found in section \ref{section2} and are given by:
\begin{align}
&R_{0}(z_1,z_1,z_2) = \frac{-1}{2z_1^6z_2^3} \\[7pt]
&r_{\frac{1}{2},0}(z_1) =\frac{-1}{4z_1^2}+\frac{1}{z_1}H_{2z_1} \\[7pt]
  &r_{1,0}(z_1) =\frac{-f_{1,0}(z_1)}{2y(z_1)}+\sum_{k=1}^{\infty}\frac{2(-1)^k \left(\frac{k}{2}\right)^2f_{1,0}(\frac{k}{2})}{z_1(z_1-\frac{k}{2})(z_1+\frac{k}{2})} \\[7pt]
  &r_{\frac{1}{2},0}(z_1,z_2)=\frac{-f_{\frac{1}{2},0}(z_1,z_2)}{2y(z_1)}+\sum_{k=1}^{\infty}\frac{2(-1)^k\left(\frac{k}{2}\right)^2f_{\frac 12,0}(\frac{k}{2},z_2)}{z_1(z_1-\frac{k}{2})(z_1+\frac{k}{2})} \\[7pt]
 & + \frac{1}{y(z_2)}r_{\frac{1}{2},0}(z_2)\left(R_{0}(z_1,z_2)+\frac{1}{(z_1^2-z_2^2)^2}\right)-\frac{1}{2z_1(z_2^2-z_1^2)}\left(\partial_{z_2}\frac{r_{\frac{1}{2},0}(z_2)}{y(z_2)}\right)
 \nonumber
\end{align}
with 
\begin{align}
& f_{\frac{1}{2},0}(z_1,z_2) =\frac{1}{2z_1}\partial_{z_1} R_{0}(z_1,z_2) +2r_{\frac{1}{2},0}(z_1)\left(R_{0}(z_1,z_2)+\frac{1}{(z_1^2-z_2^2)^2}\right)\\[7pt]
  &f_{1,0}(z_1) = \frac{1}{2z_1}\partial_{z_1}r_{\frac{1}{2},0}(z_1) + R_{0}(z_1,z_1)+r_{\frac{1}{2},0}(z_1)^2 
\end{align}
Using the above information the residues at zero are straightforward to compute using a computer algebra system:
\begin{equation}
    \begin{aligned}
        &v_{1,0}(b_1,b_2) \supset \underset{z_2=0}{\Res}\hspace{.1cm}\underset{z_1=0}{\Res} \frac{-4z_1z_2f_{1,0}(z_1,z_2)e^{b_1 z_2}e^{b_2 z_2}}{2b_1 b_2y(z_1)}+ \order{b_1^{-1}}+\order{b_2^{-1}} \\[8pt] 
  & =\frac{1}{12} b_1 \left(b_2^3+4 \pi ^2 b_2+96 \zeta (3)\right)+\frac{7 b_1^4}{96}+\left(\frac{7 b_2^2}{48}+\frac{5 \pi ^2}{6}\right) b_1^2+\frac{45 b_2^4+720 \pi ^2 b_2^2+2416 \pi ^4}{1440} \\ 
  \label{t2a}
  & +\underset{k_1=1}{\overset{\infty }{\sum }}\left(\frac{-64 (-1)^{k_1} H_{k_1}}{k_1^3}+\frac{(-1)^{k_1} \left(32 \left(H_{k_1}\right){}^2+32 \psi ^{(1)}\left(k_1+1\right)\right)}{ k_1^2}\right). 
    \end{aligned}
\end{equation}
The sums can be evaluated in terms of alternating multi zeta values,
\begin{equation}
    \begin{aligned}
    \label{A32}
       &\underset{k_1=1}{\overset{\infty }{\sum }}\left(\frac{-64 (-1)^{k_1} H_{k_1}}{k_1^3}+\frac{(-1)^{k_1} \left(32 \left(H_{k_1}\right){}^2+32 \psi ^{(1)}\left(k_1+1\right)\right)}{ k_1^2}\right) \\
      & =\underset{k_1=1}{\overset{\infty }{\sum }}\left(\frac{-64 (-1)^{k_1} H_{k_1}}{k_1^3}+\frac{32(-1)^{k_1} \left( \left(H_{k_1}\right){}^2-H^{(2)}_{k_1}+ \zeta(2)\right)}{ k_1^2}\right) \\
      &= 64\left(\frac{1}{2}\zeta(2)\zeta(\overline{2})-\zeta(\overline{4})+\zeta(\overline{2},1,1)\right)=\frac{8\pi^4}{45}+64\zeta(\overline{2},1,1)
    \end{aligned}
\end{equation}
We note that:
\begin{align}
\label{A33}
    \zeta(\overline{2},1,1) = -\frac{1}{2} \zeta(2,1,1)=-\frac{1}{2}\zeta(4),
\end{align}
so that
\begin{align}
\label{A34}
    \frac{8\pi^4}{45}+64\zeta(\overline{2},1,1) = -\frac{8\pi^4}{45}.
\end{align}
Therefore the contribution from the residues at zero is
\begin{equation}
    \begin{aligned}
         &v_{1,0}(b_1,b_2) \supset \underset{z_2=0}{\Res}\hspace{.1cm}\underset{z_1=0}{\Res} \frac{-4z_1z_2f_{1,0}(z_1,z_2)e^{b_1 z_2}e^{b_2 z_2}}{2b_1 b_2y(z_1)}+ \order{b_1^{-1}}+\order{b_2^{-1}} \\[8pt] 
  & =\frac{1}{12} b_1 \left(b_2^3+4 \pi ^2 b_2+96 \zeta (3)\right)+\frac{7 b_1^4}{96}+\left(\frac{7 b_2^2}{48}+\frac{5 \pi ^2}{6}\right) b_1^2+\frac{b_2^4}{32}+\frac{b_2^2\pi^2}{2}+\frac{3\pi^4}{2} . 
  \label{T1}
    \end{aligned}
\end{equation}
The next contribution we compute is,
\begin{equation}
\begin{aligned}
  v_{1,0}(b_1,b_2) &\supset \sum_{k=1}^{\infty}\underset{z_2=-k/2}{\Res}\hspace{.1cm}\underset{z_1=0}{\Res} -\frac{4z_1z_2f_{1,0}(z_1,z_2)e^{b_1 z_2}e^{b_2 z_2}}{2b_1 b_2y(z_1)} \\[8pt] 
  & =8b_1 \left(\text{Li}_3\left(-e^{-\frac{b_2}{2}}\right)+ \text{Li}_3\left(e^{-\frac{b_2}{2}}\right)\right)+\sum_{k_2=1}^{\infty}\underset{k_1 \neq k_2}{\sum_{k=1}^{\infty}}\frac{16 b_1 \left((-1)^{k_1}+(-1)^{k_2}\right) e^{-\frac{1}{2} b_2 k_1}}{k_1 \left(k_1-k_2\right) \left(k_1+k_2\right)} \\
  &=8b_1 \left(\text{Li}_3\left(-e^{-\frac{b_2}{2}}\right)+ \text{Li}_3\left(e^{-\frac{b_2}{2}}\right)\right)-16b_1\left(\text{Li}_{3}\left(-e^{-\frac{1}{2}b_2}\right)+\frac{1}{2}\text{Li}_{3}\left(e^{-\frac{1}{2}b_2}\right)\right) \\[8pt]
  &=-8b_1\text{Li}_{3}\left(-e^{-\frac{1}{2}b_2}\right)
  \label{T2}
\end{aligned}
\end{equation}
The next term is,
\begin{align}
    v_{1,0}(b_1,b_2) &\supset \sum_{k=1}^{\infty}\underset{z_2=0}{\Res}\hspace{.1cm}\underset{z_1=\frac{-k}{2}}{\Res} -\frac{4z_1z_2f_{1,0}(z_1,z_2)e^{b_1 z_2}e^{b_2 z_2}}{2b_1 b_2y(z_1)} \\[8pt] \nonumber
  & =-8 b_2 \text{Li}_3\left(-e^{-\frac{b_1}{2}}\right).
  \label{T3}
\end{align}
Next,
\begin{equation}
\begin{aligned}
 &v_{1,0}(b_1,b_2) \supset  \sum_{k_2=1}^{\infty}\underset{k_1\neq k_2}{\sum_{k_1=1}^{\infty}}\underset{z_2=-\frac{k_2}{2}}{\Res}\hspace{.1cm}\underset{z_1=-\frac{k_1}{2}}{\Res}-\frac{4z_1z_2f_{1,0}(z_1,z_2)e^{b_1 z_2}e^{b_2 z_2}}{2b_1 b_2y(z_1)} \\[8pt] 
 &= \sum_{k_1 \neq k_2} 16 (-1)^{k_1+k_2} e^{-\frac{1}{2}\left(b_1k_1+b_2k_2\right)}\left(\frac{-1}{k_1^2(k_1+k_2)^2}+\frac{1}{k_1^2(k_1-k_2)^2}+\frac{1}{k_1^3(k_1-k_2)}+\frac{-1}{k_1^3(k_1+k_2)}\right)\\[8pt] 
 &-32 \left(\text{Li}_4\left(e^{-\frac{b_1}{2}-\frac{b_2}{2}}\right)+\text{Li}_3\left(-e^{-\frac{b_1}{2}}\right) \log \left(e^{-\frac{b_2}{2}}+1\right)\right) \\[8pt] 
 &=\frac{16}{4}\text{Li}_4\left(e^{-\frac{1}{2}\left(b_1+b_2\right)}\right)-16\text{Li}_{2,2}\left(-e^{-\frac{1}{2}b_2},e^{-\frac{1}{2}\left(b_1-b_2\right)}\right)+16\text{Li}_{2,2}\left(e^{-\frac{1}{2}\left(b_1+b_2\right)},-e^{\frac{1}{2}b_2}\right) \\[8pt] 
 &+16\text{Li}_2\left(e^{-\frac{1}{2}\left(b_1+b_2\right)}\right)\text{Li}_2\left(-e^{-\frac{1}{2}b_2}\right) -16\text{Li}_3\left(e^{-\frac{1}{2}\left(b_1+b_2\right)}\right)\text{Li}_1\left(-e^{-\frac{1}{2}b_2}\right)\\[8pt]
 &+16\text{Li}_{3,1}\left(e^{-\frac{1}{2}\left(b_1+b_2\right)},-e^{\frac{1}{2}b_2}\right)+\frac{16}{2}\text{Li}_4\left(e^{-\frac{1}{2}\left(b_1+b_2\right)}\right)-16\text{Li}_{1,3}\left(-e^{-\frac{1}{2}b_2},e^{-\frac{1}{2}\left(b_1-b_2\right)}\right)\\[8pt]
 &-32 \left(\text{Li}_4\left(e^{-\frac{b_1}{2}-\frac{b_2}{2}}\right)-\text{Li}_3\left(-e^{-\frac{b_1}{2}}\right) \text{Li}_1\left(-e^{-\frac{b_2}{2}}\right)\right) \\[8pt] 
&=12\text{Li}_4\left(e^{-\frac{1}{2}\left(b_1+b_2\right)}\right)-16\text{Li}_{2,2}\left(-e^{-\frac{1}{2}b_2},e^{-\frac{1}{2}\left(b_1-b_2\right)}\right)+16\text{Li}_{2,2}\left(e^{-\frac{1}{2}\left(b_1+b_2\right)},-e^{\frac{1}{2}b_2}\right) \\[8pt] 
 &+16\text{Li}_{2,2}\left(e^{-\frac{1}{2}\left(b_1+b_2\right)},-e^{-\frac{1}{2}b_2}\right)+16\text{Li}_{2,2}\left(-e^{-\frac{1}{2}b_2},e^{-\frac{1}{2}\left(b_1+b_2\right)}\right)-16\text{Li}_{3,1}\left(e^{-\frac{1}{2}\left(b_1+b_2\right)},-e^{-\frac{1}{2}b_2}\right) \\[8pt]
 &-16\text{Li}_{1,3}\left(-e^{-\frac{1}{2}b_2},e^{-\frac{1}{2}\left(b_1+b_2\right)}\right)+16\text{Li}_{3,1}\left(e^{-\frac{1}{2}\left(b_1+b_2\right)},-e^{\frac{1}{2}b_2}\right)-16\text{Li}_{1,3}\left(-e^{-\frac{1}{2}b_2},e^{-\frac{1}{2}\left(b_1-b_2\right)}\right)\\[8pt]
 &+32\text{Li}_{3,1}\left(-e^{-\frac{b_1}{2}},-e^{-\frac{b_2}{2}}\right) +32\text{Li}_{1,3}\left(-e^{-\frac{b_2}{2}},-e^{-\frac{b_1}{2}}\right)  
 \label{T4}
\end{aligned}
\end{equation}
where in the last line we stuffled the result.
Next,
\begin{equation}
\begin{aligned}
   & v_{1,0}(b_1,b_2) \supset \sum_{k=1}^{\infty}\underset{z_2=-\frac{k}{2}}{\Res}\hspace{.1cm}\underset{z_1=-\frac{k}{2}}{\Res}\frac{-4z_1z_2f_{1,0}(z_1,z_2)e^{b_1 z_2}e^{b_2 z_2}}{2b_1 b_2y(z_1)} \\[8pt]
   & =\sum_{k_1=1}^{\infty}\left(-\frac{8\left( b_1+b_2\right) e^{-\frac{1}{2} \left(b_1+b_2\right) k_1} H_{k_1-1}}{k_1^2}-\frac{32 e^{-\frac{1}{2} \left(b_1+b_2\right) k_1} H_{k_1-1}}{k_1^3}-\frac{16 e^{-\frac{1}{2} \left(b_1+b_2\right) k_1} H_{k_1-1}^{(2)}}{k_1^2}\right)\\[8pt] 
   &+\left(2 b_1^2+2b_2^2+2b_1b_2+32\zeta(2)\right) \text{Li}_2\left(e^{-\frac{b_1}{2}-\frac{b_2}{2}}\right)+\left(8 b_1 +8b_2\right)\text{Li}_3\left(e^{-\frac{b_1}{2}-\frac{b_2}{2}}\right)+20 \text{Li}_4\left(e^{-\frac{b_1}{2}-\frac{b_2}{2}}\right)\\[8pt] 
   &=\left(2 b_1^2+2b_2^2+2b_1b_2+32\zeta(2)\right) \text{Li}_2\left(e^{-\frac{b_1}{2}-\frac{b_2}{2}}\right)+\left(8 b_1 +8b_2\right)\left(\text{Li}_3\left(e^{-\frac{b_1}{2}-\frac{b_2}{2}}\right)-\text{Li}_{2,1}\left(e^{-\frac{b_1}{2}-\frac{b_2}{2}},1\right)\right) \\[8pt]  
   &+20 \text{Li}_4\left(e^{-\frac{b_1}{2}-\frac{b_2}{2}}\right)-32\text{Li}_{3,1}\left(e^{-\frac{b_1}{2}-\frac{b_2}{2}},1\right)-16\text{Li}_{2,2}\left(e^{-\frac{b_1}{2}-\frac{b_2}{2}},1\right) \\[8pt]
   &=\left(2 b_1^2+2b_2^2+2b_1b_2\right) \text{Li}_2\left(e^{-\frac{b_1}{2}-\frac{b_2}{2}}\right)+\left(8 b_1 +8b_2\right)\left(\text{Li}_3\left(e^{-\frac{b_1}{2}-\frac{b_2}{2}}\right)-\text{Li}_{2,1}\left(e^{-\frac{b_1}{2}-\frac{b_2}{2}},1\right)\right) \\[8pt]  
   &+52 \text{Li}_4\left(e^{-\frac{b_1}{2}-\frac{b_2}{2}}\right)-32\text{Li}_{3,1}\left(e^{-\frac{b_1}{2}-\frac{b_2}{2}},1\right)+16\text{Li}_{2,2}\left(e^{-\frac{1}{2}\left(b_1+b_2\right)},1\right)+32\text{Li}_{2,2}\left(1,e^{-\frac{1}{2}\left(b_1+b_2\right)}\right)
   \label{T5}
\end{aligned}
\end{equation}
where in the last equality we stuffled the result.
Finally,
\begin{equation}
\begin{aligned}
     & v_{1,0}(b_1,b_2) \supset \sum_{k=1}^{\infty}\underset{z_2=-\frac{k}{2}}{\Res}\hspace{.1cm}\underset{z_1=-\frac{k}{2}}{\Res}\frac{-4z_1z_2f_{1,0}(z_1,z_2)e^{b_1 z_2}e^{b_2 z_2}}{2b_1 b_2y(z_1)} \\[8pt]
     &=\sum_{k=1}^{\infty}\left(-\frac{8 \left(b_1-b_2\right) e^{-\frac{1}{2} \left(b_1-b_2\right) k_1} H_{k_1-1}}{k_1^2}-\frac{32 e^{-\frac{1}{2} \left(b_1-b_2\right) k_1} H_{k_1-1}}{k_1^3}-\frac{16 e^{-\frac{1}{2} \left(b_1-b_2\right) k_1} H_{k_1-1}^{(2)}}{k_1^2}\right)\\[8pt] 
     &+\left(2 b_1^2 -2b_2b_1 +16\zeta(2)\right) \text{Li}_2\left(e^{\frac{1}{2} \left(b_2-b_1\right)}\right)+8 b_1 \text{Li}_3\left(e^{\frac{1}{2} \left(b_2-b_1\right)}\right)\\[8pt] 
     &=\left(2 b_1^2 -2b_2b_1 +16\zeta(2)\right) \text{Li}_2\left(e^{\frac{1}{2} \left(b_2-b_1\right)}\right)+8 b_1 \text{Li}_3\left(e^{\frac{1}{2} \left(b_2-b_1\right)}\right)+ 8\left(b_2-b_1\right)\text{Li}_{2,1}\left(e^{\frac{1}{2}\left(b_2-b_1\right)},1\right)\\[8pt]&
    -32\text{Li}_{3,1}\left(e^{\frac{1}{2}\left(b_2-b_1\right)},1\right)-16\text{Li}_{2,2}\left(e^{\frac{1}{2}\left(b_2-b_1\right)},1\right)\\[8pt]
   & =\left(2 b_1^2 -2b_2b_1 \right) \text{Li}_2\left(e^{\frac{1}{2} \left(b_2-b_1\right)}\right)+8 b_1 \text{Li}_3\left(e^{\frac{1}{2} \left(b_2-b_1\right)}\right)+16 \text{Li}_4\left(e^{\frac{1}{2} \left(b_2-b_1\right)}\right)\\[8pt] 
    &+16\text{Li}_{2,2}\left(1,e^{\frac{1}{2}\left(b_2-b_1\right)}\right)+ 8\left(b_2-b_1\right)\text{Li}_{2,1}\left(e^{\frac{1}{2}\left(b_2-b_1\right)},1\right)-32\text{Li}_{3,1}\left(e^{\frac{1}{2}\left(b_2-b_1\right)},1\right).
    \label{T6}
\end{aligned}
\end{equation}
The full result is then found by adding together~\eqref{T1} -~\eqref{T6}:
\begin{equation}
\begin{aligned}
     v_{1,0}(b_1,b_2) &=\frac{1}{12} b_1 \left(b_2^3+4 \pi ^2 b_2+96 \zeta (3)\right)+\frac{7 b_1^4}{96}+\left(\frac{7 b_2^2}{48}+\frac{5 \pi ^2}{6}\right) b_1^2+\frac{b_2^4}{32}+\frac{b_2^2\pi^2}{2}+\frac{3\pi^4}{2} \\[8pt]
    &-8b_1\text{Li}_{3}\left(-e^{-\frac{1}{2}b_2}\right) -8b_2\text{Li}_{3}\left(e^{-\frac{1}{2}b_1}\right)+\left(2 b_1^2+2b_2^2+2b_1b_2\right) \text{Li}_2\left(e^{-\frac{b_1}{2}-\frac{b_2}{2}}\right) \\[8pt]
    &+8 b_1 \text{Li}_3\left(e^{\frac{1}{2} \left(b_2-b_1\right)}\right)+8\left( b_1 +b_2\right)\left(\text{Li}_3\left(e^{-\frac{b_1}{2}-\frac{b_2}{2}}\right)-\text{Li}_{2,1}\left(e^{-\frac{b_1}{2}-\frac{b_2}{2}},1\right)\right)\\[8pt]
    &+\left(2 b_1^2 -2b_2b_1 \right) \text{Li}_2\left(e^{\frac{1}{2} \left(b_2-b_1\right)}\right)+8 \left(b_2-b_1\right)\text{Li}_{2,1}\left(e^{\frac{1}{2}\left(b_2-b_1\right)},1\right)+16 \text{Li}_4\left(e^{\frac{1}{2} \left(b_2-b_1\right)}\right)\\[8pt]
    &+64\text{Li}_4\left(e^{-\frac{1}{2}\left(b_1+b_2\right)}\right)-32\text{Li}_{3,1}\left(e^{-\frac{1}{2}\left(b_1+b_2\right)},1\right)-32\text{Li}_{3,1}\left(e^{-\frac{1}{2}\left(b_1-b_2\right)},1\right)\\[8pt]
    &+16\text{Li}_{2,2}\left(e^{-\frac{1}{2}\left(b_1+b_2\right)},1\right)+32\text{Li}_{2,2}\left(1,e^{-\frac{1}{2}\left(b_1+b_2\right)}\right)+16\text{Li}_{2,2}\left(1,e^{-\frac{1}{2}\left(b_1-b_2\right)}\right) \\[8pt]
 &  -16\text{Li}_{2,2}\left(-e^{-\frac{1}{2}b_2},e^{-\frac{1}{2}\left(b_1-b_2\right)}\right)+16\text{Li}_{2,2}\left(e^{-\frac{1}{2}\left(b_1+b_2\right)},-e^{\frac{1}{2}b_2}\right) \\[8pt] 
 &+16\text{Li}_{2,2}\left(e^{-\frac{1}{2}\left(b_1+b_2\right)},-e^{-\frac{1}{2}b_2}\right)+16\text{Li}_{2,2}\left(-e^{-\frac{1}{2}b_2},e^{-\frac{1}{2}\left(b_1+b_2\right)}\right)\\[8pt]
 &-16\text{Li}_{3,1}\left(e^{-\frac{1}{2}\left(b_1+b_2\right)},-e^{-\frac{1}{2}b_2}\right) -16\text{Li}_{1,3}\left(-e^{-\frac{1}{2}b_2},e^{-\frac{1}{2}\left(b_1+b_2\right)}\right)\\[8pt]
 &+16\text{Li}_{3,1}\left(e^{-\frac{1}{2}\left(b_1+b_2\right)},-e^{\frac{1}{2}b_2}\right)-16\text{Li}_{1,3}\left(-e^{-\frac{1}{2}b_2},e^{-\frac{1}{2}\left(b_1-b_2\right)}\right)\\[8pt]
 &+32\text{Li}_{3,1}\left(-e^{-\frac{b_1}{2}},-e^{-\frac{b_2}{2}}\right) +32\text{Li}_{1,3}\left(-e^{-\frac{b_2}{2}},-e^{-\frac{b_1}{2}}\right).
\end{aligned}
\end{equation}

\section{Details of double trumpet integrations}
\label{appendixc}
\subsection{$g=\frac{1}{2}$}
\label{g1/2}
We begin by computing the $g=\frac{1}{2}$ contribution to the $\tau-$scaled SFF. The necessary volume is:
\begin{align}
   V^{(p)}_{\frac{1}{2}}(b_1,b_2) &= 4\log(p) + b_1+ 2\log(1+e^{\frac{b_2-b_1}{2}})+2\log(1+e^{\frac{-b_2-b_1}{2}})
\end{align}
which we use the $b_1 \leftrightarrow b_2 $ symmetry to write as:
\begin{align}
   V^{(p)}_{\frac{1}{2}}(b_1,b_2)= V^{(p)}_{\frac{1}{2}}(b_1,b_2)\theta(b_1-b_2) +V^{(p)}_{\frac{1}{2}}(b_2,b_1)\theta(b_2-b_1).
\end{align}
The $g=\frac{1}{2}$ contribution to the $\tau-$scaled SFF is then:
\begin{align}
  \kappa_\frac{1}{2}(t,\beta) &=  \frac{1}{t^2}\int_0^\infty \int_0^\infty b_1\dd{b_1} b_2\dd{b_2}Z^t(b_1,\beta_1) Z^t(b_2,\beta_2)V^{(p)}_{\frac{1}{2}}(b_1,b_2)\theta(b_1-b_2)\\ \nonumber
  &+\left(\beta_1 \leftrightarrow \beta_2\right) ,
\end{align}
with $\beta_1 = \beta +it$ and $\beta_2=\beta - it$. Direct integration shows the contribution from $v_{\frac{1}{2},1}(b_1,b_2)= 4$ is $\order{t^{-\frac 12}}$ and does not need to be considered. The contribution from the Airy term, $v_{\frac 12,0}(b_1,b_2) \supset b_1$, is:
\begin{equation}
    \begin{aligned}
        \kappa_\frac{1}{2}(t,\beta) &\supset  \frac{1}{t^2}\int_0^\infty \int_0^\infty b_1\dd{b_1} b_2\dd{b_2}Z^t(b_1,\beta_1) Z^t(b_2,\beta_2)b_1\theta(b_1-b_2)\\ 
  &+\left(\beta_1 \leftrightarrow \beta_2\right) \\[7pt]
  &=\frac{-1}{\sqrt{2\pi \beta}}+\order{t^{-\frac{1}{2}}}.
  \label{7}
    \end{aligned}
\end{equation}
The contribution from $v_{\frac 12,0}(b_1,b_2) \supset 2\log(1+e^{\frac{b_2-b_1}{2}})$ is slightly more complicated but can be computed as follows:

\begin{align}
\label{k1/23}
    &\kappa_\frac{1}{2}(t,\beta) \supset  \frac{1}{t^2}\int_0^\infty \int_0^\infty b_1\dd{b_1} b_2\dd{b_2}Z^t(b_1,\beta_1) Z^t(b_2,\beta_2)2\log(1+e^{\frac{b_2-b_1}{2}})\theta(b_1-b_2)\\ \nonumber
  &+\left(\beta_1 \leftrightarrow \beta_2\right) \\[7pt]
  &= \sum_{k=1}^{\infty}\frac{2(-1)^{k+1}}{k} \frac{1}{t^2}\int_0^\infty \int_0^\infty b_1\dd{b_1} b_2\dd{b_2}Z^t(b_1,\beta_1) Z^t(b_2,\beta_2)e^{\frac{k(b_2-b_1)}{2}}\theta(b_1-b_2) \\ \nonumber
  &+\left(\beta_1 \leftrightarrow \beta_2\right) \\[7pt] 
  \label{5}
  & =\sum_{k=1}^{\infty}\frac{(-1)^k}{k}\frac{1}{t^2}\int_0^{\infty}db_2 \frac{b_2 \beta_1 }{2\pi\sqrt{\beta_2} \sqrt{\beta_1}
  } e^{-\frac{b_2^2 \left(\beta _1+\beta _2\right)}{4 \beta _1 \beta _2}} \\[7pt] \nonumber
  &\times \left(\sqrt{\pi \beta_1 } k e^{\frac{\left(b_2+\beta _1 k\right){}^2}{4 \beta _1}} \text{erfc}\left(\frac{\sqrt{\beta_2}b_2}{2\sqrt{\beta_1}\sqrt{ \beta_2}}+\frac{\beta _1 k}{2 \sqrt{\beta _1}}\right)-2\right) + \left(\beta_2 \leftrightarrow \beta_1\right).
\end{align}
The integral can be solved by making the substitution $b_2 \to 2\sqrt{\beta_1}\sqrt{\beta_2}b_2$ and then using the asymtotic expansion of the erfc$(x)$:
\begin{align}
  \text{erfc}(x) \cong \frac{e^{-x^2}}{x\sqrt{\pi}}\sum_{n=0}^{\infty}(-1)^n\frac{(2n-1)!!}{(2x^2)^n}.
\end{align}
Only the $n=0$ term in the asymptotic expansion will produce a contribution in the long time limit. The solution to~\eqref{5} is then:
\begin{align}
  \frac{-2}{\sqrt{2\beta \pi}}\sum_{k=1}^{\infty}(-1)^k\left(1-k\sqrt{\frac{\beta \pi}{2}}e^{\frac{\beta k^2}{2}}\text{erfc}\left(\frac{\sqrt{\beta}k}{\sqrt{2}}\right)\right)+\order{t^{-\frac 1 2}}
  \label{6}
\end{align}
By following similar steps it can be shown the contribution from $v_{\frac 12,0}(b_1,b_2) \supset 2\log(1+e^{-\frac{b_2+b_1}{2}})$ is subleading in the long time limit. The result quoted in the main text~\eqref{k1/22} is found by combining~\eqref{7} and~\eqref{6}.
\subsection{$g=1$}
\label{g1}
The coefficient of $\log(p)^2$ in the $g=1$ contribution can be computed from the following subset of terms, 
\begin{equation}
    v_{1,2}(b_1,b_2) \supset 2b_1^2+2b_2^2
    \label{b10}
\end{equation}
which would give the following contributions to the SFF, after introducing theta functions in the same way that was done for $g=\frac{1}{2}$:
\begin{equation}
\begin{aligned}
    &\frac{1}{t^3}\int_0^\infty \int_0^\infty b_1\dd{b_1} b_2\dd{b_2}Z^t(b_1,\beta_1) Z^t(b_2,\beta_2)v_{1,2}(b_1,b_2)\theta(b_1-b_2)\\[8pt]
  &+\left(\beta_1 \leftrightarrow \beta_2\right)+\order{t^{-\frac 1 2}} \\[8pt]
  &= 0 +\order{t^{-\frac 1 2}}.
  \end{aligned}
\end{equation}
Only the terms in~\eqref{b10} give terms that are not subleading in $t$, however, integrating ~\eqref{b10} against the double trumpet gives exactly zero.

The coefficient of $\log(p)$ in the $g=1$ contribution can be computed from the following subset of terms:
\begin{equation}
  v_{1,1}(b_1,b_2)  \supset \frac{1}{3} \left(2 b_1^3+3 b_2^2 b_1+b_2^3\right) +8\left(b_1-b_2\right) \text{Li}_2\left(e^{\frac{1}{2} \left(b_2-b_1\right)}\right).
  \label{b12}
\end{equation}
The evaluation of the term with $\text{Li}_2\left(e^{\frac{1}{2} \left(b_2-b_1\right)}\right)$ consists of expanding the function in a power series\footnote{The expansion in a power series is only valid after the introduction of theta functions.} and following the same steps taken in~\eqref{k1/23}-\eqref{6}, except more terms in the asymptotic expansion of erfc$(x)$ need to be considered. We find the contribution from the generic term $b_1 e^{\frac{k(b_2-b_1)}{2}}$ after integrating against the double trumpet to be:
\begin{align}
  \frac{k }{\beta \pi}\left(1-\frac{\beta k^2}{2}e^{\frac{\beta k^2}{2}}E_1\left(\frac{\beta k^2}{2}\right)\right) +\order{t^{-\frac{1}{2}}}
  \label{8}
\end{align} 
It turns out the contribution from $b_2 e^{\frac{k(b_2-b_1)}{2}}$ is the exact same as~\eqref{8}. Therefore,
\begin{equation}
\begin{aligned}
    &\frac{1}{t^3}\int_0^\infty \int_0^\infty b_1\dd{b_1} b_2\dd{b_2}Z^t(b_1,\beta_1) Z^t(b_2,\beta_2)8(b_1-b_2)\text{Li}_2\left(e^{\frac{1}{2} \left(b_2-b_1\right)}\right)\theta(b_1-b_2)\\[8pt]
  &+\left(\beta_1 \leftrightarrow \beta_2\right)+\order{t^{-\frac 1 2}} \\[8pt]
  &= 0 +\order{t^{-\frac 1 2}}.
  \end{aligned}
\end{equation}
The contribution from the polynomial term in~\eqref{b12} is also exactly zero, so that,
\begin{equation}
\begin{aligned}
    &\frac{1}{t^3}\int_0^\infty \int_0^\infty b_1\dd{b_1} b_2\dd{b_2}Z^t(b_1,\beta_1) Z^t(b_2,\beta_2)v_{1,1}(b_1,b_2)\theta(b_1-b_2)\\[8pt]
  &+\left(\beta_1 \leftrightarrow \beta_2\right)+\order{t^{-\frac 1 2}} \\[8pt]
  &= 0 +\order{t^{-\frac 1 2}}.
  \end{aligned}
\end{equation}
The finite part of the $g=1$ contribution can be computed from the following subset of terms of the unorientable volume:
\begin{equation}
    \begin{aligned}
      v_{1,0}(b_1,b_2) &\supset 
\frac{1}{12} b_1 \left(b_2^3+4 \pi ^2 b_2\right)+\frac{7 b_1^4}{96}+\left(\frac{7 b_2^2}{48}+\frac{5 \pi ^2}{6}\right) b_1^2+\frac{b_2^4}{32}+\frac{b_2^2\pi^2}{2} \\[8pt]
&+\left(2 b_1^2 -2b_2b_1\right) \text{Li}_2\left(e^{\frac{1}{2} \left(b_2-b_1\right)}\right)+8b_1 \text{Li}_3\left(e^{\frac{1}{2} \left(b_2-b_1\right)}\right)\\[8pt]
&+8(b_2-b_1) \text{Li}_{2,1}\left(e^{\frac{1}{2} \left(b_2-b_1\right)},1\right).
   \label{universal1}
    \end{aligned}
\end{equation}
We again introduce theta functions so that the contribution to the $\tau-$scaled SFF is:
\begin{equation}
    \begin{aligned}
        \label{k11}
   \kappa_1(t,\beta) &=  \frac{1}{t^3}\int_0^\infty \int_0^\infty b_1\dd{b_1} b_2\dd{b_2}Z^t(b_1,\beta_1) Z^t(b_2,\beta_2)v_{1,0}(b_1,b_2)\theta(b_1-b_2)\\
  &+\left(\beta_1 \leftrightarrow \beta_2\right)+\order{t^{-\frac 1 2}},
    \end{aligned}
\end{equation}
The subset of terms:
\begin{align}
 v_{1,0}(b_1,b_2)\supset \frac{1}{12} b_1 \left(b_2^3+4 \pi ^2 b_2\right)+\frac{7 b_1^4}{96}+\left(\frac{7 b_2^2}{48}+\frac{5 \pi ^2}{6}\right) b_1^2+\frac{b_2^4}{32}+\frac{b_2^2\pi^2}{2},
\end{align}
can be directly integrated to give:
\begin{align}
  \kappa_1(t,\beta) \supset \frac{-2\pi }{3\beta} + \frac{1}{\pi}\left(\frac{-10}{3}+\log(\frac{2t}{\beta})\right) +\order{t^{-\frac{1}{2}}}.
  \label{14}
\end{align}
Based on the previous discussion of~\eqref{8}, the contribution of 
\begin{equation}
   8 (b_2-b_1) \text{Li}_{2,1}\left(e^{\frac{1}{2} \left(b_2-b_1\right)},1\right) +8b_1\text{Li}_3\left(e^{\frac{1}{2} \left(b_2-b_1\right)}\right) 
\end{equation}
is 
\begin{equation}
\begin{aligned}
   \kappa_1(t,\beta) &\supset  0 +  \sum_{k=1}^{\infty}\frac{8 }{k^2\beta \pi}\left(1-\frac{\beta k^2}{2}e^{\frac{\beta k^2}{2}}E_1\left(\frac{\beta k^2}{2}\right)\right) +\order{t^{-\frac{1}{2}}} \\[8pt]
   & =\frac{4\pi}{3\beta}-\frac{4}{\pi}\sum_{k=1}^{\infty}e^{\frac{\beta k^2}{2}}E_1\left(\frac{\beta k^2}{2}\right),
   \label{b21}
\end{aligned}
\end{equation}
and the contribution of the term,
\begin{align}
 \left(2 b_1^2 -2b_2b_1\right) \text{Li}_2\left(e^{\frac{1}{2} \left(b_2-b_1\right)}\right),
\end{align} is,
\begin{align}
  \kappa_{1}(t,\beta) \supset \frac{-2\pi}{3\beta} +\frac{4}{\pi}\sum_{k=1}^{\infty} \left(  e^{\frac{\beta k^2}{2}} \left(\frac{\beta k^2}{2}+\frac{3}{2}\right) E_1\left(\frac{k^2 \beta}{2} \right)-1\right)+\order{t^{-\frac{1}{2}}},
  \label{13}
\end{align}
Adding the contributions from~\eqref{b21}, \eqref{13}, and~\eqref{14}:
\begin{equation}
    \begin{aligned}
        \kappa_1(t,\beta)&= \frac{-2\pi }{3\beta} + \frac{1}{\pi}\left(\frac{-10}{3}+\log(\frac{2t}{\beta})\right) + \frac{-2 \pi}{3\beta} +\frac{4\pi}{3\beta}\\[8pt] 
 &+ \frac{4}{\pi}\sum_{k=1}^{\infty}\left(  e^{\frac{\beta k^2}{2}} \left(\frac{\beta k^2}{2}+\frac{3}{2}\right) E_1\left(\frac{k^2 \beta}{2} \right)-1 -e^{\frac{\beta k^2}{2}} E_1\left(\frac{k^2 \beta}{2} \right)\right) \\[8pt]
  &  +\order{t^{-\frac{1}{2}}}\\[8pt] 
  &=\frac{1}{\pi}\left(\frac{-10}{3}+\log(\frac{2t}{\beta})\right)+\frac{4}{\pi}\sum_{k=1}^{\infty}\left(-1+\frac{1}{2}\left(1+\beta k^2\right)e^{\frac{\beta k^2}{2}}E_1\left(\frac{\beta k^2}{2}\right)\right) \\[8pt]
  &+\order{t^{-\frac{1}{2}}}
    \end{aligned}
\end{equation}
which is the result quoted in the main text~\eqref{k10}.
\section{Details of RMT integrations}
\label{appendixd}
The $\tau-$scaled SFF from universal RMT follows from putting the form factor~\eqref{eq:cf} into~\eqref{eq18}:
\begin{equation}
  \begin{aligned}
  \kappa^{\text{GOE}}(\tau,\beta)&=
    \int_{0}^{E^{*}} \dd{E} e^{-2\beta E}\qty[2\rho_0(E)-\frac{\tau}{2\pi}\log\qty(\frac{\tau}{\pi}+\rho_0(E))+\frac{\tau}{2\pi}\log(\frac{\tau}{\pi}-\rho_0(E))]\\[10pt]
    & +\int_{E^{*}}^{\infty}\dd{E}e^{-2\beta E}\qty[\frac{\tau}{\pi}-\frac{\tau}{2\pi}\log\qty(1+\frac{\tau}{\pi\rho_0(E)})],
\end{aligned}\label{eq:SFF_t_GOE_universal}
\end{equation}
with 
\begin{align}
  \rho_0(E^{*}) = \frac{\tau}{2\pi}.
\end{align}
It can be conveniently written as:
\begin{align}
 \kappa^{\text{GOE}}\left(\tau,\beta\right)=2\kappa^{\text{GUE}}(\tau,\beta)+\chi(\tau,\beta),
\end{align}
where we have defined:
\begin{align}
\nonumber
  & \chi(\tau,\beta) \coloneqq\\
  & -\frac{\tau}{2\pi}\qty[\int_{0}^{\infty} \dd{E} e^{-2\beta E}\log\qty(1+\frac{\tau}{\pi\rho_0(E)})-\int_{0}^{E^{*}}\dd{E}e^{-2\beta E}\log\qty(-1+\frac{\tau}{\pi\rho_0(E)})].
  \label{chi1}
\end{align}
and $\kappa^{\text{GUE}}(\tau,\beta)$ was computed in \cite{Saad2022}:
\begin{equation}
    \begin{aligned}
        2\kappa^{\text{GUE}}(\tau,\beta) &= \frac{2e^{\frac{\pi^2}{2\beta}}}{16 \sqrt{2\pi}\beta^{3/2}}\left(\text{Erf}\left(\frac{\frac{\beta}{\pi}\text{arcsinh}(2\pi \tau)+\pi}{\sqrt{2\beta}}\right)+\text{Erf}\left(\frac{\frac{\beta}{\pi}\text{arcsinh}(2\pi \tau)-\pi}{\sqrt{2\beta}}\right)\right) \\
  &=\frac{\tau}{2\pi \beta}-\frac{\tau^3}{3\pi} +\order{\tau^4}.
  \label{C3}
    \end{aligned}
\end{equation}
Focusing on the bracketed part of~\eqref{chi1}, the logarithms can be split up as:
\begin{align}
  & \int_0^{\infty}\dd{E} e^{-2 \beta E} \log\qty(\frac{\tau}{\pi}+ \rho_0(E))-\int_0^{E^{*}}dE e^{-2\beta E}\log\qty(\frac{\tau}{\pi} - \rho_0(E)) - \int_{E^{*}}^{\infty}dE e^{-2\beta E}\log\qty(\rho_0(E)).
\end{align}
Each integral can be integrated by parts and the sum of the boundary terms vanishes.
The remaining integrals are then:
\begin{equation}
  \frac{1}{2\beta}\qty[\int_0^{\infty}e^{-2 \beta E} \frac{\rho'_0(E)dE}{\left(\frac{\tau}{\pi}+ \rho_0(E)\right)}+\int_0^{E^{*}} e^{-2\beta E}\frac{\rho_0'(E)dE}{\left(\frac{\tau}{\pi} - \rho_0(E)\right)} - \int_{E^{*}}^{\infty} e^{-2\beta E}\frac{\rho_0'(E)dE}{\rho_0(E)}].
\end{equation}
Introducing the JT gravity leading order energy density~\eqref{rho} this can be written as:

\begin{equation}
    \begin{aligned}
        &\frac{1}{2\beta}\qty[\int_0^{\infty}\dd{x}\left( e^{-\frac{2 \beta}{4\pi^2}\text{arcsinh}\left(x-\tau\right)^2}\frac{1}{(x+\tau)} -e^{-\frac{2 \beta}{4\pi^2}\text{arcsinh} (x+\tau)^2}\frac{1}{(x+\tau)}\right)] =\\[8pt]
  &\frac{1}{2\beta}\sum_{n=1}^{\infty}\frac{(-2\tau)^n}{n!}\int_{\tau}^{\infty}dx\frac{\frac{d^n}{dx^n}e^{-\frac{2\beta}{4\pi^2}\left(\text{arcsinh}(2\pi x)\right)^2}}{x}.
  \label{eq:J8} 
    \end{aligned}
\end{equation}
To get to the last line we used the translation operator and then made the substitution: $x \to x+\tau$. $\chi(\tau,\beta)$ is then given by:
\begin{align}
    \chi(\tau,\beta) =
\frac{-\tau}{4\pi\beta}\sum_{n=1}^{\infty}\frac{(-2\tau)^n}{n!}\int_{\tau}^{\infty}dx\frac{\frac{d^n}{dx^n}e^{-\frac{2\beta}{4\pi^2}\left(\text{arcsinh}(2\pi x)\right)^2}}{x}.
\label{chi2}
\end{align}
The integral can be solved perturbatively in $\tau$. To compute the integral to $\order{\tau^4}$, only the $n=1$ and $n=2$ terms need to be considered. The $n=1$ term of~\eqref{chi2} gives:
\begin{equation}
    \begin{aligned}
        &\frac{2\tau^2}{4\pi \beta}\int^{\infty}_{\frac{\sqrt{2\beta}}{2\pi}\text{arcsinh}(2\pi \tau)}dy\frac{-4\pi y e^{-y^2}}{\sinh(\frac{2\pi y}{\sqrt{2\beta}})} \\[8pt]
 &=\frac{-2\tau^2}{4\pi \beta}\int^{\infty}_{\frac{\sqrt{2\beta}}{2\pi}\text{arcsinh}(2\pi \tau)}dye^{-y^2}4 y\left(\frac{\sqrt{2\beta}}{2y}+\frac{4 y}{\sqrt{2\beta}}\sum_{k=1}^{\infty}\frac{(-1)^k }{\frac{4y^2}{2\beta}+k^2}\right) \\[8pt]
 &=  -\frac{\tau^2}{\sqrt{2\pi \beta}}+ \frac{2\tau^3}{\pi}-\frac{2\tau^2}{\sqrt{2\pi \beta}} \sum_{k=1}^{\infty} (-1)^k \left(1-k\sqrt{\frac{\beta \pi}{2}} e^{\frac{\beta k^2}{2}} \text{erfc}\left(k\sqrt{\frac{\beta}{2}}\right)\right) +\order{\tau^4}.
 \label{c2}
    \end{aligned}
\end{equation}
To get the second line we used the series expansion:
\begin{align}
  \frac{\pi}{\sinh(\pi x)}=\frac{1}{x}+\sum_{k=1}^{\infty}\frac{2x (-1)^k}{x^2+k^2},
\end{align}
and to get the third line we used the integral definitions of the incomplete gamma function and erfc$(x)$. The $n=2$ term of~\eqref{chi2} is:
\begin{equation}
\begin{aligned}
  &\frac{-2\tau^3}{\beta}\int_{\frac{\sqrt{2\beta}}{2\pi}\text{archsinh}(2\pi \tau)}^{\infty}dy  e^{-y^2} \left(2\sqrt{2\beta} \left(2 y^2-1\right) \csch\left(\frac{2\sqrt{2} \pi y}{\sqrt{\beta}}\right)+2 \pi y\text{sech}^2\left(\frac{\sqrt{2} \pi y}{\sqrt{\beta}}\right)\right)=\\
   &=\frac{-2\tau^3}{\beta}\int_{\frac{\sqrt{2\beta}}{2\pi}\text{archsinh}(2\pi \tau)}^{\infty}dy  e^{-y^2} \left(2\sqrt{2\beta} \left(2 y^2-1\right) \csch\left(\frac{2\sqrt{2} \pi y}{\sqrt{\beta}}\right)+\sqrt{2\beta} \left(2 y^2-1\right) \text{tanh}\left(\frac{\sqrt{2} \pi y}{\sqrt{\beta}}\right)\right) \\ 
   &=\frac{-2\tau^3}{\beta}\int_{\frac{\sqrt{2\beta}}{2\pi}\text{archsinh}(2\pi \tau)}^{\infty}dy  e^{-y^2} \left(\sqrt{2\beta} \left(2 y^2-1\right) \coth\left(\frac{\sqrt{2} \pi y}{\sqrt{\beta}}\right)\right) .
\end{aligned}
\end{equation}
In the second line we integrated by parts. Using the series expansion for $\coth(x)$ and performing the integrals results in:
\begin{align}
   &\frac{-2\tau^3}{\beta \pi}\int_{\frac{\sqrt{2\beta}}{2\pi}\text{archsinh}(2\pi \tau)}^{\infty}dy  e^{-y^2} \sqrt{2 \beta}(2y^2-1)\left(\frac{\sqrt{\beta}}{\sqrt{2}y}+\frac{2\sqrt{2}y}{\sqrt{\beta}}\sum_{k=1}^{\infty}\left(\frac{1}{\frac{2y^2}{\beta}+k^2}\right)\right) \nonumber\\[10pt] 
   &=\frac{-2\tau^3}{\pi} -\frac{\tau^3}{\pi}\left(\gamma + \log(2\beta \tau^2)\right)+\frac{-4 \tau^3}{\pi}\sum_{k=1}^{\infty}
   \left(1-\frac{1}{2}e^{\frac{\beta k^2}{2}}(1+\beta k^2)E_1\left(\frac{\beta k^2}{2}\right)\right)\nonumber \\
   &+\order{\tau^4}.
   \label{C1}
\end{align}
The result quoted in the main text~\eqref{GOE} is found by combining~\eqref{C3},~\eqref{c2}, and ~\eqref{C1}.

\bibliography{lib}

\providecommand{\href}[2]{#2}\begingroup\raggedright\begin{thebibliography}{10}

\bibitem{Jackiw1985}
R.~Jackiw, \emph{{Lower dimensional gravity}}, \href{https://doi.org/10.1016/0550-3213(85)90448-1}{\emph{Nuclear Physics, Section B} {\bfseries 252} (1985) 343}.

\bibitem{Teitelboim1983}
C.~Teitelboim, \emph{{Gravitation and hamiltonian structure in two spacetime dimensions}}, \href{https://doi.org/10.1016/0370-2693(83)90012-6}{\emph{Physics Letters B} {\bfseries 126} (1983) 41}.

\bibitem{Saad2019}
P.~Saad, S.H.~Shenker and D.~Stanford, \emph{{JT gravity as a matrix integral}},  \href{https://arxiv.org/abs/1903.11115}{{\ttfamily 1903.11115}}.

\bibitem{Kitaev2016_1}
A.~Kitaev, ``A simple model of quantum holography (part 1).'' talk at KITP \url{http://online.kitp.ucsb.edu/online/entangled15/kitaev/}, April 7, 2015.

\bibitem{Kitaev2016_2}
A.~Kitaev, ``{A simple model of quantum holography (part 2)}.'' talk at KITP \url{http://online.kitp.ucsb.edu/online/entangled15/kitaev2/}, May 27, 2015.

\bibitem{Maldacena2016b}
J.~Maldacena and D.~Stanford, \emph{{Remarks on the Sachdev-Ye-Kitaev model}}, \href{https://doi.org/10.1103/PhysRevD.94.106002}{\emph{Physical Review D} {\bfseries 94} (2016) } [\href{https://arxiv.org/abs/1604.07818}{{\ttfamily 1604.07818}}].

\bibitem{Jensen2016}
K.~Jensen, \emph{{Chaos in $AdS_2$ Holography}}, \href{https://doi.org/10.1103/PhysRevLett.117.111601}{\emph{Physical Review Letters} {\bfseries 117} (2016) 111601} [\href{https://arxiv.org/abs/1605.06098}{{\ttfamily 1605.06098}}].

\bibitem{Mertens2023}
T.G.~Mertens and G.J.~Turiaci, \emph{Solvable models of quantum black holes: a review on jackiw--teitelboim gravity}, \href{https://doi.org/10.1007/s41114-023-00046-1}{\emph{Living Reviews in Relativity} {\bfseries 26} (2023) 4}.

\bibitem{Stanford2019}
D.~Stanford and E.~Witten, \emph{{JT} gravity and the ensembles of random matrix theory}, \href{https://doi.org/10.4310/ATMP.2020.v24.n6.a4}{\emph{Advances in Theoretical and Mathematical Physics} {\bfseries 24} 1475}.

\bibitem{Dyson1962a}
F.J.~Dyson, \emph{{Statistical Theory of the Energy Levels of Complex Systems. I}}, \href{https://doi.org/10.1063/1.1703773}{\emph{Journal of Mathematical Physics} {\bfseries 3} (1962) 140}.

\bibitem{Altland1997}
A.~Altland and M.R.~Zirnbauer, \emph{{Nonstandard symmetry classes in mesoscopic normal-superconducting hybrid structures}}, \href{https://doi.org/10.1103/PhysRevB.55.1142}{\emph{Physical Review B} {\bfseries 55} (1997) 1142}.

\bibitem{Stanford2023}
D.~Stanford, \emph{{A Mirzakhani recursion for non-orientable surfaces}},  \href{https://arxiv.org/abs/arXiv:2303.04049}{{\ttfamily arXiv:2303.04049}}.

\bibitem{Maldacena2016}
J.~Maldacena, D.~Stanford and Z.~Yang, \emph{{Conformal symmetry and its breaking in two-dimensional nearly anti-de Sitter space}}, \href{https://doi.org/10.1093/ptep/ptw124}{\emph{Progress of Theoretical and Experimental Physics} {\bfseries 2016} (2016) } [\href{https://arxiv.org/abs/1606.01857}{{\ttfamily 1606.01857}}].

\bibitem{Bohigas1984}
O.~Bohigas, M.J.~Giannoni and C.~Schmit, \emph{{Characterization of Chaotic Quantum Spectra and Universality of Level Fluctuation Laws}}, \href{https://doi.org/10.1103/PhysRevLett.52.1}{\emph{Physical Review Letters} {\bfseries 52} (1984) 1}.

\bibitem{Saad2018}
P.~Saad, S.H.~Shenker and D.~Stanford, \emph{{A semiclassical ramp in SYK and in gravity}},  \href{https://arxiv.org/abs/1806.06840}{{\ttfamily 1806.06840}}.

\bibitem{Cotler2017}
J.S.~Cotler, G.~Gur-Ari, M.~Hanada, J.~Polchinski, P.~Saad, S.H.~Shenker et~al., \emph{{Black holes and random matrices}}, \href{https://doi.org/10.1007/JHEP05(2017)118}{\emph{Journal of High Energy Physics} {\bfseries 2017} (2017) } [\href{https://arxiv.org/abs/1611.04650}{{\ttfamily 1611.04650}}].

\bibitem{Altland2020}
A.~Altland and J.~Sonner, \emph{{Late time physics of holographic quantum chaos}}, \href{https://doi.org/10.21468/SCIPOSTPHYS.11.2.034}{\emph{SciPost Physics} {\bfseries 11} (2021) } [\href{https://arxiv.org/abs/2008.02271}{{\ttfamily 2008.02271}}].

\bibitem{Saad2022}
P.~Saad, D.~Stanford, Z.~Yang and S.~Yao, \emph{{A convergent genus expansion for the plateau}}, \href{https://doi.org/https://doi.org/10.1007/JHEP09(2024)033}{\emph{Journal of High Energy Physics} {\bfseries 2024} (2024) } [\href{https://arxiv.org/abs/2210.11565}{{\ttfamily 2210.11565}}].

\bibitem{Weber2024}
T.~Weber, J.~Tall, F.~Haneder, J.D.~Urbina and K.~Richter, \emph{{Unorientable topological gravity and orthogonal random matrix universality}}, \href{https://doi.org/https://doi.org/10.1007/JHEP07(2024)267}{\emph{Journal of High Energy Physics} {\bfseries 2024} (2024) } [\href{https://arxiv.org/abs/2405.17177}{{\ttfamily 2405.17177}}].

\bibitem{Norbury2008}
P.~Norbury, \emph{{Lengths of geodesics on non-orientable hyperbolic surfaces}}, \href{https://doi.org/10.1007/s10711-008-9251-3}{\emph{Geometriae Dedicata} {\bfseries 134} (2008) 153}.

\bibitem{Mirzakhani2006}
M.~Mirzakhani, \emph{{Weil-Petersson volumes and intersection theory on the moduli space of curves}}, \href{https://doi.org/10.1090/S0894-0347-06-00526-1}{\emph{Journal of the American Mathematical Society} {\bfseries 20} (2006) 1}.

\bibitem{Mirzakhani2007}
M.~Mirzakhani, \emph{{Simple geodesics and Weil-Petersson volumes of moduli spaces of bordered Riemann surfaces}}, \href{https://doi.org/10.1007/s00222-006-0013-2}{\emph{Inventiones Mathematicae} {\bfseries 167} (2007) 179}.

\bibitem{Gendulphe2017}
M.~Gendulphe, \emph{{What's wrong with the growth of simple closed geodesics on nonorientable hyperbolic surfaces}},  \href{https://arxiv.org/abs/1706.08798}{{\ttfamily 1706.08798}}.

\bibitem{Eynard2007}
B.~Eynard and N.~Orantin, \emph{{Weil-Petersson volume of moduli spaces, Mirzakhani's recursion and matrix models}},  \href{https://arxiv.org/abs/0705.3600}{{\ttfamily 0705.3600}}.

\bibitem{Weber2022}
T.~Weber, F.~Haneder, K.~Richter and J.D.~Urbina, \emph{{Constraining Weil-Petersson volumes by universal random matrix correlations in low-dimensional quantum gravity}}, \href{https://doi.org/10.1088/1751-8121/acc8a5}{\emph{Journal of Physics A: Mathematical and Theoretical} {\bfseries 56} (2023) } [\href{https://arxiv.org/abs/2208.13802}{{\ttfamily 2208.13802}}].

\bibitem{Blommaert2022}
A.~Blommaert, J.~Kruthoff and S.~Yao, \emph{{An integrable road to a perturbative plateau}}, \href{https://doi.org/10.1007/JHEP04(2023)048}{\emph{Journal of High Energy Physics} {\bfseries 2023} (2023) } [\href{https://arxiv.org/abs/2208.13795}{{\ttfamily 2208.13795}}].

\bibitem{Okuyama2021}
K.~Okuyama and K.~Sakai, \emph{{'t Hooft expansion of multi-boundary correlators in 2D topological gravity}}, \href{https://doi.org/10.1093/ptep/ptab090}{\emph{Progress of Theoretical and Experimental Physics} {\bfseries 2021} (2021) 083B03}.

\bibitem{Okuyama2023}
K.~Okuyama and K.~Sakai, \emph{{Spectral form factor in the $\tau$-scaling limit}}, \href{https://doi.org/10.1007/JHEP04(2023)123}{\emph{Journal of High Energy Physics} {\bfseries 2023} (2023) } [\href{https://arxiv.org/abs/2301.04773}{{\ttfamily 2301.04773}}].

\bibitem{Mehta2004}
M.L.~Mehta, \emph{{Random Matrices}}, ISSN, Elsevier Science (2004).

\bibitem{Haake2010}
F.~Haake, \emph{{Quantum signatures of chaos}}, Springer series in synergetics, Springer, Berlin [u.a.] (2010).

\bibitem{Eynard2016}
B.~Eynard, \emph{{Counting Surfaces}}, vol.~70 of \emph{Progress in Mathematical Physics} (2016).

\bibitem{Artemev2024}
A.~Artemev and I.~Chaban, \emph{$(2p+1)$ minimal string and intersection theory 1},  \href{https://arxiv.org/abs/2403.02305}{{\ttfamily 2403.02305}}.

\bibitem{Mertens2021}
T.G.~Mertens and G.J.~Turiaci, \emph{{Liouville quantum gravity {--} holography, JT and matrices}}, \href{https://doi.org/10.1007/JHEP01(2021)073}{\emph{Journal of High Energy Physics} {\bfseries 2021} (2021) } [\href{https://arxiv.org/abs/2006.07072}{{\ttfamily 2006.07072}}].

\bibitem{Eynard2018}
B.~Eynard, T.~Kimura and S.~Ribault, \emph{{Random matrices}},  \href{https://arxiv.org/abs/1510.04430}{{\ttfamily 1510.04430}}.

\bibitem{Eynard2004a}
B.~Eynard, \emph{{Topological expansion for the 1-hermitian matrix model correlation functions}}, \href{https://doi.org/10.1088/1126-6708/2004/11/031}{\emph{Journal of High Energy Physics} {\bfseries 8} (2004) 845}.

\bibitem{Waldschmidt2002}
M.~Waldschmidt, \emph{Multiple polylogarithms: An introduction},  in \emph{Number Theory and Discrete Mathematics}, (Basel), pp.~1--12, Birkh{\"a}user Basel, 2002.

\bibitem{Gil2023}
J.~Gil and J.~Fresan, \emph{Multiple zeta values: from numbers to motives}, {\emph{in press} (2017) }.

\bibitem{Zhao2016}
J.~Zhao, \emph{Multiple zeta functions, multiple polylogarithms and their special values}, \href{https://doi.org/10.1142/9634}{\emph{World Scientific} (2016) }.

\bibitem{Witten1991a}
E.~Witten, \emph{{On quantum gauge theories in two dimensions}}, \href{https://doi.org/10.1007/BF02100009}{\emph{Communications in Mathematical Physics} {\bfseries 141} (1991) 153}.

\bibitem{maximon2003}
L.C.~Maximon, \emph{The dilogarithm function for complex argument}, \href{https://doi.org/10.1098/rspa.2003.1156}{\emph{Proceedings of the Royal Society of London. Series A} {\bfseries 459} (2003) 2807}.

\bibitem{Duhr2019}
C.~Duhr and F.~Dulat, \emph{{PolyLogTools \textemdash{} polylogs for the masses}}, \href{https://doi.org/10.1007/JHEP08(2019)135}{\emph{JHEP} {\bfseries 08} (2019) 135} [\href{https://arxiv.org/abs/1904.07279}{{\ttfamily 1904.07279}}].

\bibitem{Zagier1994}
D.~Zagier, \emph{Values of zeta functions and their applications},  in \emph{First European Congress of Mathematics Paris, July 6--10, 1992: Vol. II: Invited Lectures (Part 2)}, A.~Joseph, F.~Mignot, F.~Murat, B.~Prum and R.~Rentschler, eds., (Basel), pp.~497--512, Birkh{\"a}user Basel (1994), \href{https://doi.org/10.1007/978-3-0348-9112-7_23}{DOI}.

\bibitem{Duhr2014}
C.~Duhr, \emph{{Mathematical aspects of scattering amplitudes}},  in \emph{{Theoretical Advanced Study Institute in Elementary Particle Physics}: {Journeys Through the Precision Frontier: Amplitudes for Colliders}}, pp.~419--476, 2015, \href{https://doi.org/10.1142/9789814678766_0010}{DOI} [\href{https://arxiv.org/abs/1411.7538}{{\ttfamily 1411.7538}}].

\bibitem{Schlotterer:2012}
O.~Schlotterer and S.~Stieberger, \emph{{Motivic Multiple Zeta Values and Superstring Amplitudes}}, \href{https://doi.org/10.1088/1751-8113/46/47/475401}{\emph{J. Phys. A} {\bfseries 46} (2013) 475401} [\href{https://arxiv.org/abs/1205.1516}{{\ttfamily 1205.1516}}].

\bibitem{Mafra:2022}
C.R.~Mafra and O.~Schlotterer, \emph{{Tree-level amplitudes from the pure spinor superstring}}, \href{https://doi.org/10.1016/j.physrep.2023.04.001}{\emph{Phys. Rept.} {\bfseries 1020} (2023) 1} [\href{https://arxiv.org/abs/2210.14241}{{\ttfamily 2210.14241}}].

\bibitem{Goncharov1998}
A.B.~Goncharov, \emph{{Multiple polylogarithms, cyclotomy and modular complexes}}, \href{https://doi.org/10.4310/MRL.1998.v5.n4.a7}{\emph{Math. Res. Lett.} {\bfseries 5} (1998) 497} [\href{https://arxiv.org/abs/1105.2076}{{\ttfamily 1105.2076}}].

\bibitem{Goncharov2001}
A.B.~Goncharov, \emph{Multiple polylogarithms and mixed tate motives},  \href{https://arxiv.org/abs/math/0103059}{{\ttfamily math/0103059}}.

\bibitem{brown2012}
F.~Brown, \emph{Mixed tate motives over $\mathbb{Z}$}, \href{https://doi.org/https://doi.org/10.4007/annals.2012.175.2.10}{\emph{Annals of Mathematics} {\bfseries 172} (2012) }.

\bibitem{Gnutzmann2004}
S.~Gnutzmann and B.~Seif, \emph{{Universal spectral statistics in Wigner-Dyson, chiral, and Andreev star graphs. I. Construction and numerical results}}, \href{https://doi.org/10.1103/PhysRevE.69.056219}{\emph{Physical Review E} {\bfseries 69} (2004) 16}.

\bibitem{Griguolo2024}
L.~Griguolo, J.~Papalini, L.~Russo and D.~Seminara, \emph{{The resurgence of the plateau in supersymmetric $ \mathcal{N} = 1$ Jackiw-Teitelboim gravity}}, \href{https://doi.org/10.1007/JHEP06(2024)168}{\emph{Journal of High Energy Physics} {\bfseries 2024} (2024) 168} [\href{https://arxiv.org/abs/2310.06768}{{\ttfamily 2310.06768}}].

\bibitem{Turiaci2023}
G.J.~Turiaci and E.~Witten, \emph{{$ \mathcal{N} $ = 2 JT supergravity and matrix models}}, \href{https://doi.org/10.1007/JHEP12(2023)003}{\emph{JHEP} {\bfseries 12} (2023) 003} [\href{https://arxiv.org/abs/2305.19438}{{\ttfamily 2305.19438}}].

\end{thebibliography}\endgroup

\end{document}